\documentclass{aa}
\usepackage[varg]{txfonts}
\usepackage{graphicx}
\usepackage{natbib}
\usepackage{color}
\usepackage{sidecap}

\usepackage[hidelinks=true]{hyperref}
\definecolor{dark-red}{rgb}{0.4,0.15,0.15}
\definecolor{dark-blue}{rgb}{0.1,0.1,0.6}
\definecolor{medium-blue}{rgb}{0,0,0.5}
\hypersetup{
    colorlinks, linkcolor={dark-red},
    citecolor={dark-blue}, urlcolor={medium-blue}
}
\pdfoutput=1

\begin{document}

\title{Sunspot splitting triggering an eruptive flare}

\author{Rohan E. Louis\inst{1}
  \and Klaus G. Puschmann\inst{1} \and Bernhard Kliem\inst{2,3,4} \and Horst Balthasar\inst{1} \and Carsten Denker\inst{1}}

\offprints{Rohan E. Louis $[$rlouis@aip.de$]$}

\institute{Leibniz-Institut f\"ur Astrophysik Potsdam (AIP),
	  An der Sternwarte 16, 14482 Potsdam, Germany \and
	  Yunnan Astronomical Observatory, Chinese Academy of
          Sciences, Kunming 650011, China \and
          Institute of Physics and Astronomy, University of Potsdam, 
          D-14476 Potsdam, Germany \and
          Mullard Space Science Laboratory, University College London,
          Holmbury St.\ Mary, Dorking, Surrey, RH5 6NT, UK}
\date{Received 15 January 2013 / Accepted ...}

\abstract
   {}
   {We investigate how the splitting of the leading sunspot and associated flux emergence and cancellation in 
active region NOAA 11515 caused an eruptive M5.6 flare on 2012 July 2.}
   {Continuum intensity, line-of-sight magnetogram, and dopplergram data of the Helioseismic and Magnetic Imager were employed to analyse 
the photospheric evolution. Filtergrams in H$\alpha$ and \ion{He}{I}~10830~\AA~ of the Chromospheric Telescope at the Observatorio del Teide, 
Tenerife, track the evolution of the flare. The corresponding coronal conditions were derived from 171~{\AA} and 304~{\AA} images of the 
Atmospheric Imaging Assembly. Local correlation tracking was utilized to determine shear flows.}
   {Emerging flux formed a neutral line ahead of the leading sunspot and new satellite spots. The sunspot splitting caused a long-lasting 
flow towards this neutral line, where a filament formed. Further flux emergence, partly of mixed polarity, as well as episodes of flux 
cancellation occurred repeatedly at the neutral line. Following a nearby C-class precursor flare with signs of interaction with the 
filament, the filament erupted nearly simultaneously with the onset of the M5.6 flare and evolved into a coronal mass 
ejection. The sunspot stretched without forming a light bridge, splitting unusually fast (within about a day, complete $\approx6$~hours 
after the eruption) in two nearly equal parts. The front part separated strongly from the active region to approach the neighbouring 
active region where all its coronal magnetic connections were rooted. It also rotated rapidly (by $4.9^\circ$~hr$^{-1}$) and caused 
significant shear flows at its edge.}
   {The eruption resulted from a complex sequence of processes in the (sub-) photosphere and corona. The persistent flows towards the neutral 
line likely caused the formation of a flux rope that held the filament. These flows, their associated flux cancellation, the emerging 
flux, and the precursor flare all contributed to the destabilization of the flux rope. We interpret the sunspot splitting as the separation 
of two flux bundles differently rooted in the convection zone and only temporarily joined in the spot. This explains the rotation as the continued 
rise of the separating flux, and it implies that at least this part of the sunspot was still connected to its roots deep in the convection zone.} 

\maketitle

\section{Introduction}
\label{intro}
Solar flares are energetic phenomena involving plasma heating, particle acceleration, and the release of electromagnetic 
energy spanning the range from X-rays to radio wavelengths. Eruptive flares are part of a coronal disruption that leads to 
a coronal mass ejection (CME) and also often includes the eruption of a filament or prominence. Such disruptions, also referred to as solar
eruptions, originate in regions of a highly non-potential coronal field. They have a sudden onset and develop on much shorter time scales than the 
evolutionary time scales of the photosphere on the spatial scales involved. Thus, they include the gradual storage of free magnetic energy 
in the coronal field by changes in the photospheric boundary and the sudden loss of equilibrium, or onset of instability, at a critical 
point in the evolution of the coronal field \citep{forbes10}. The resulting release of free magnetic energy returns the field to a more 
potential, less stressed configuration. 

Eruptions occur invariably in highly sheared coronal field above 
photospheric polarity inversion (neutral) lines, especially if there is also a strong gradient of the magnetic field 
\citep{1968SoPh....5..187M,1990ApJS...73..159H,2006ApJ...649..490W}. Such regions show a filament channel in the chromosphere 
and often contain a filament or prominence in the corona above \citep{1998SoPh..182..107M}. Although the complete magnetic 
structure of filaments remains a mystery, the widely adopted assumption of a weakly twisted magnetic flux rope holding the cool material is 
consistent with the majority of filament properties, especially with the observed, typically inverse field direction in their bottom part 
\citep{2010SSRv..151..333M}. The formation and instability of a flux rope is a key element of storage-and-release eruption models 
\citep{1995ApJ...446..377F, 2003ApJ...595.1231A, 2006ApJ...641..577M, Kliem&Torok2006}. The resulting rise of the flux rope forms a 
vertical current sheet underneath, where flare reconnection is triggered \citep{2000JGR...105.2375L}.

The changes in the magnetic field that precede and lead to such eruptions are complex and can vary greatly from event to 
event. They may involve the emergence, shearing and/or twisting, and cancellation of flux in the photosphere, as well as small-scale 
precursor activities in the corona. Various combinations of these processes have been observed, and despite intense study, a clear picture 
of their roles in triggering eruptions has not yet emerged \citep[e.g.,][]{2008SoPh..250...75Z}. By performing a detailed analysis of a 
comprehensively observed, relatively strong eruptive flare, we address the relationship between the evolution of the photospheric flux 
distribution and the corresponding evolution of the coronal field leading to the eruption. This event features the splitting of the source 
region's leading sunspot as a special driving factor and is additionally characterized by considerable complexity in the photospheric 
changes. However, we find that the evolution towards the eruptive flare can be understood in terms of the three basic photospheric 
driving processes mentioned above acting in combination. Still, the sunspot splitting was unusual, intriguing on its own. We briefly 
characterize the basic photospheric driving processes in the following.

Flares often occur in emerging flux regions, where newly emerged fields appear in a region of pre-existing flux 
\cite[e.g.,][]{1972SoPh...25..141R,1982AdSpR...2...39M,1993A&A...271..292D,1995JGR...100.3355F,2000PASJ...52..465L, 
2004ApJ...616..578S, 2009AdSpR..43..739S, 2012ApJ...748...77S}. Many authors suggest that the emergence leads to the 
formation of a magnetic flux rope in the corona, either by bodily emergence \citep{1996SoPh..167..217L, 2005ApJ...622.1275L} or by 
reconnection within an emerged magnetic arcade \citep{2004ApJ...610..588M, 2010A&A...514A..56A}. See \citet{2012SoPh..278...33V} for a 
discussion of these competing concepts. It has been suggested that shear flows at the neutral line are intimately connected with the 
emergence process, as the primary driver of the flux rope formation \citep{2004ApJ...610..588M}. However, it is not clear whether 
the connection is as universal as has been suggested.

Horizontal shear flows in the photosphere are often seen in any evolutionary phase of an active region, as well as between 
adjacent active regions. They can be derived from photospheric data \citep{2006ApJ...644.1278D, 2009ApJ...690.1820T, 2012ApJ...761..105L} 
or be inferred from the changing appearance of coronal loop arcades \citep{2011A&A...526A...2G}. A direct relationship with flaring activity 
has been established in these and in may other cases. It is not yet clear whether flares and CMEs are triggered in highly sheared arcade 
fields \citep{2012ApJ...760...81K}, although no large-scale instability of this configuration is known or whether the high shear first 
leads to the formation of a magnetic flux rope, whose instability then causes the eruption.

Flares can also be triggered by the rapid rotation of sunspots 
\citep{2007ApJ...662L..35Z,2009RAA.....9..596Y,2009SoPh..258..203M,2012ApJ...744...50J,2012ApJ...754...16Y,2013SoPh..286..453T}, 
which twists the coronal field, such that the helical kink instability may be triggered in cases of very strong rotation. 
Rotational motions rarely occur in isolated sunspots, so in most cases they also shear the field rooted near the periphery of the rotating spot 
\citep[e.g.,][]{2008ApJ...675.1637S}.

Finally, flux cancellation is an important mechanism for triggering solar flares \citep{1989SoPh..121..197L, 
2010A&A...521A..49S, 2011A&A...526A...2G, 2012ApJ...759..105S, 2013SoPh..283..429B}. In particular, \citet{2001ApJ...548L..99Z} 
located the initial flare brightenings and filament disturbances of a major eruptive event exactly at the sites of flux 
cancellation. The link between these phenomena is usually considered to be the formation of a flux rope, which is expected 
if the cancelling photospheric flux patches lie at the base of sheared coronal flux 
\citep{1989ApJ...343..971V, 2006ApJ...641..577M, 2010ApJ...708..314A, 2011ApJ...742L..27A}. The cancellation then involves reconnection 
of the sheared field low in the atmosphere, as observationally verified by \citet{1993SoPh..143..119W}. This produces a 
flux rope in the corona.
Flux cancellation at the neutral line is a natural result of the flux dispersal in the decay phase of active regions, so this process is 
considered to be the typical driver of eruptions in decaying active regions.

\begin{figure*}[!ht]
\centerline{
\includegraphics[angle=90,width = 1.03\textwidth]{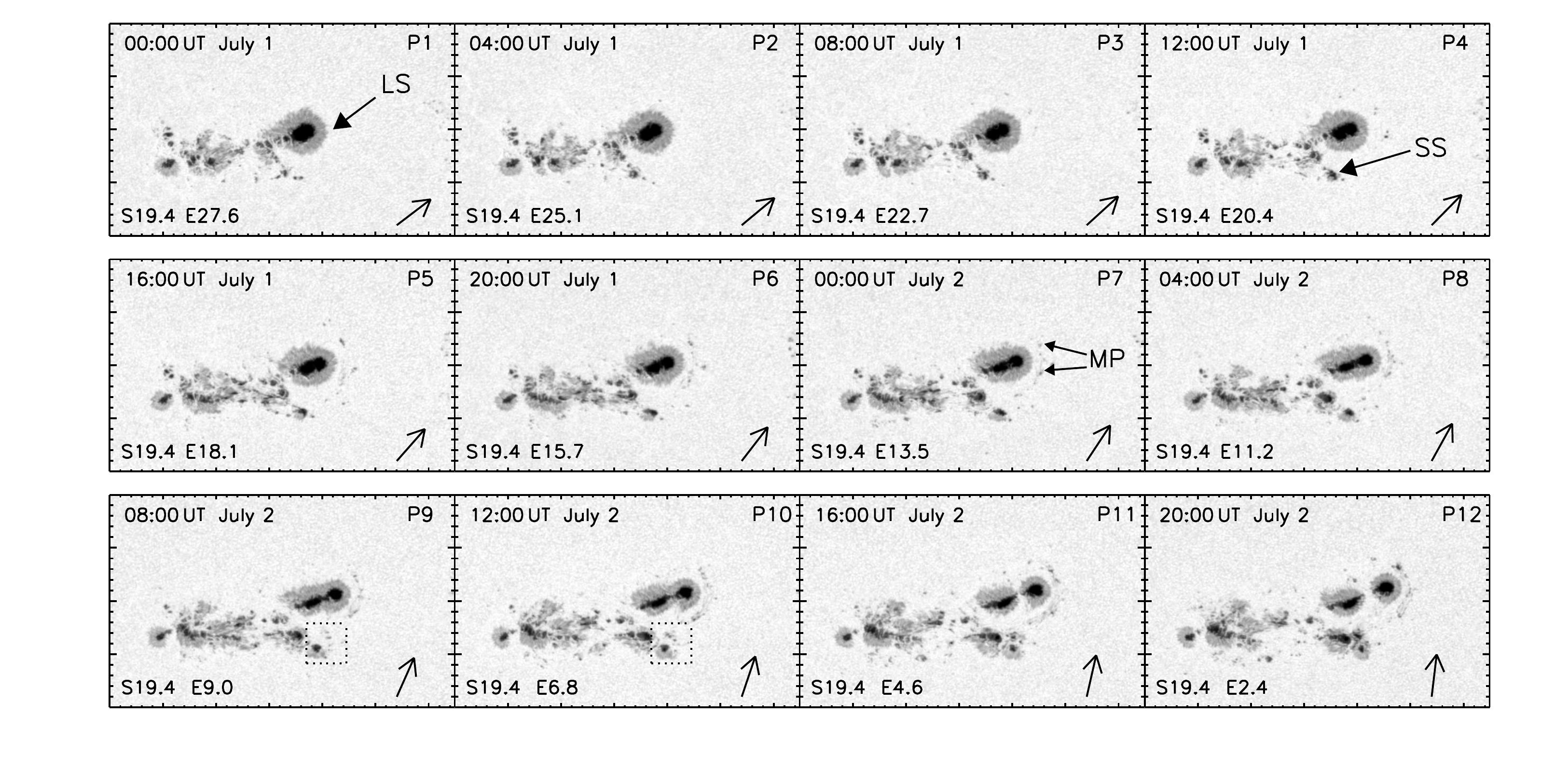}
}
\vspace{-26pt}
\centerline{
\includegraphics[angle=90,width = 1.03\textwidth]{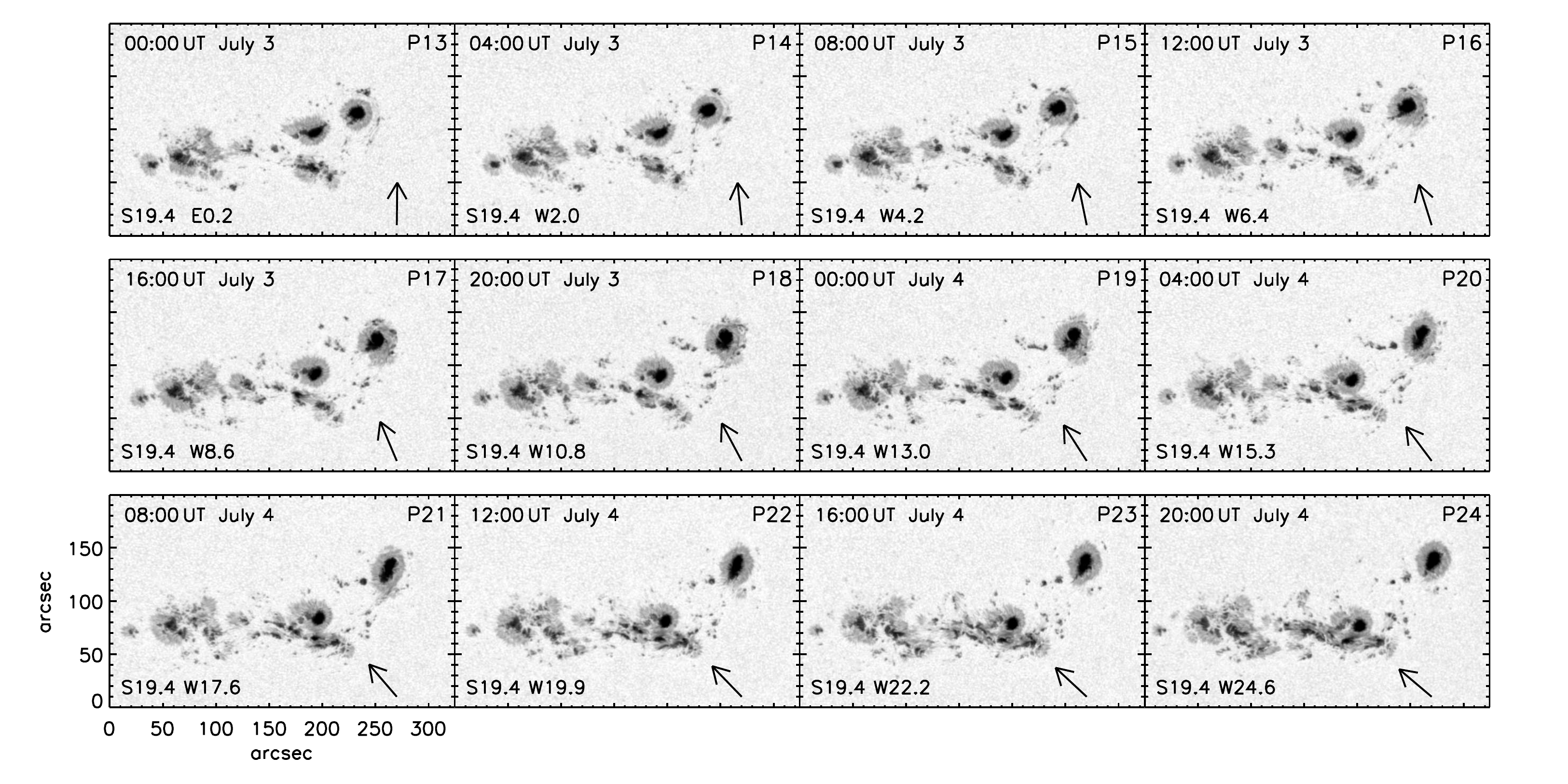}
}
\vspace{-12pt}
\caption{Time sequence of continuum images from SDO/HMI. The time and date of the observations are indicated in
the top left corner, while the position of the active region on the solar disc is shown in the bottom left corner 
of each panel. The arrow in the lower right corner of each panel points to disc centre. Solar east and north are 
to the left and top, respectively. Leading sunspot - {\bf{LS}}, the satellite group of sunspots - {\bf{SS}}, and a 
region of mixed polarity - {\bf{MP}}. The panel number is indicated on the top right corner.}
\label{figure01}
\end{figure*}

\begin{figure*}[!ht]
\centerline{
\includegraphics[angle=90,width = 1.03\textwidth]{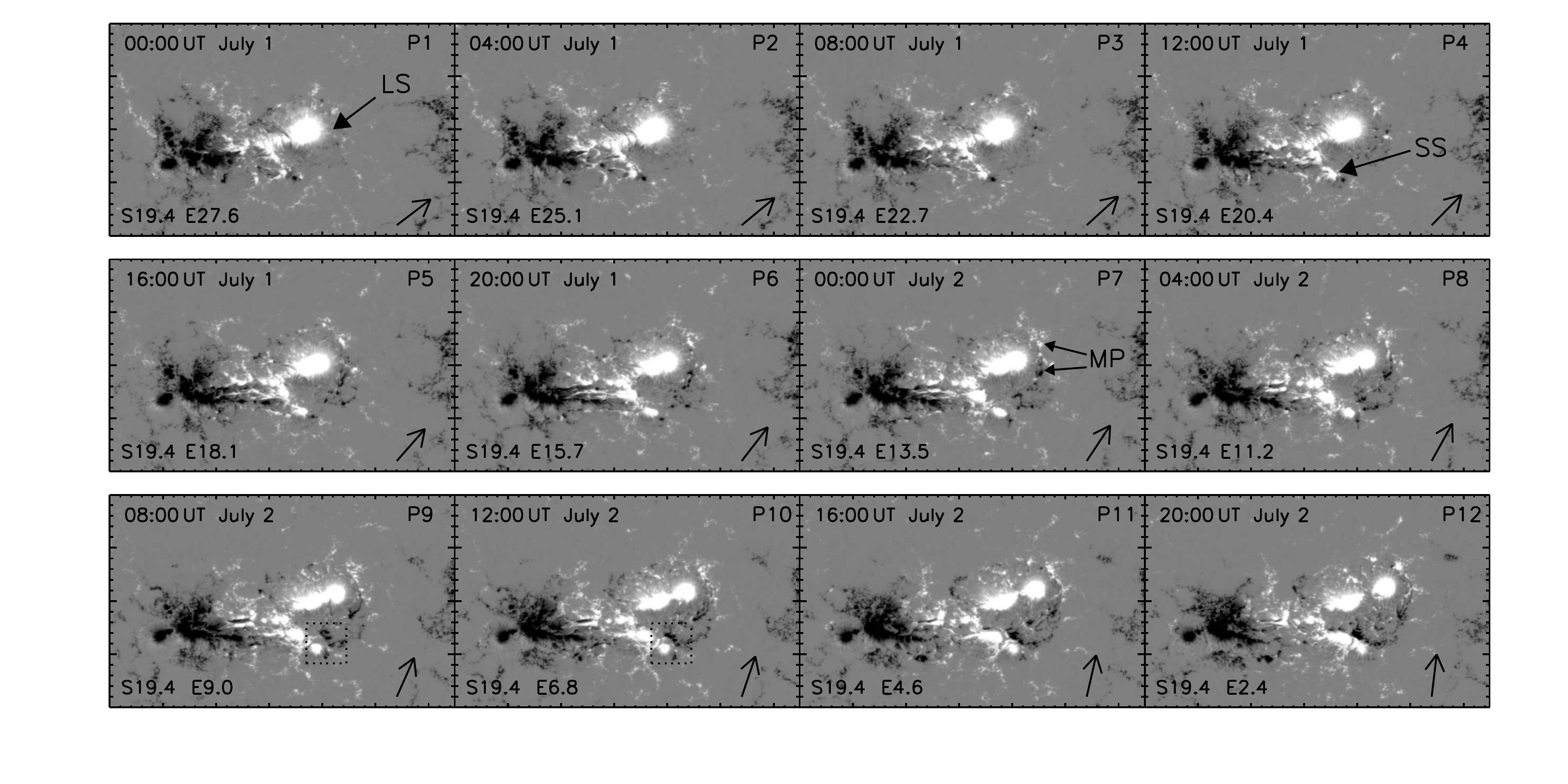}
}
\vspace{-26pt}
\centerline{
\includegraphics[angle=90,width = 1.03\textwidth]{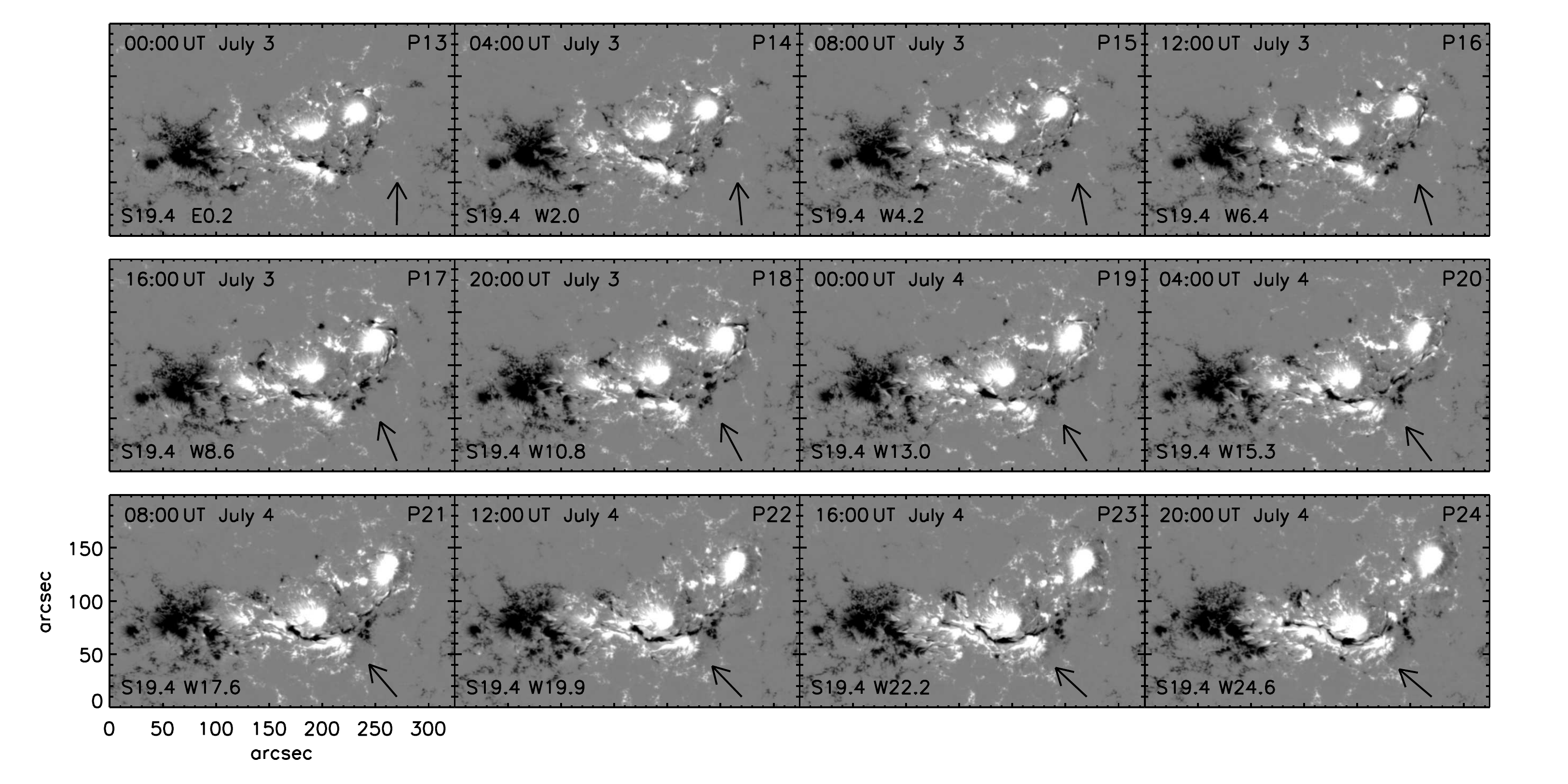}
}
\vspace{-12pt}
\caption{Same as Fig.~\ref{figure01} but for line-of-sight magnetograms.}
\label{figure02}
\end{figure*}

In this paper we analyse an eruptive M5.6 flare originating in a complex active region. We employ multi-wavelength 
observations covering the wide range of scales involved in the photosphere, chromosphere, upper transition region, 
and corona. The causal relationships between the photospheric and coronal phenomena lie in the main focus of this 
investigation, but we also address the possible origin of the sunspot splitting. The paper is organized as follows. The observations 
and procedures for data reduction are described in Sect.~\ref{data}. We describe the results of our analysis in Sect.~\ref{result}. The 
discussion and conclusion are presented in Sects.~\ref{discuss} and ~\ref{conclu}, respectively.

\section{Observations and data processing}
\label{data}
For this investigation we employed full-disc solar observations from the Solar Dynamics Observatory \citep[SDO;][]{2012SoPh..275....3P} 
which include data from the Helioseismic and Magnetic Imager \citep[HMI,][]{2012SoPh..275..229S} and the 
Atmospheric Imaging Assembly \citep[AIA,][]{2012SoPh..275...17L} for NOAA active region (AR) 11515 during 2012 July~1--4. 
The HMI data set consists of 4k~$\times$~4k-pixel images of continuum intensity, dopplergrams, and line-of-sight 
(LOS) magnetograms derived from the photospheric \ion{Fe}{i}~617.3~nm line at a cadence of 12 min and 
a spatial sampling of 0.5\arcsec~pixel$^{-1}$. The continuum images were corrected for limb darkening 
using a fifth-degree polynomial fit to the average radial profile. Projection effects were compensated for using the 
method described by \citet{1990SoPh..126...21G}. The dopplergrams and magnetograms were corrected for the $\cos\Theta$ 
factor (where $\Theta$ is the heliocentric angle) and for projection effects similar to the continuum images. All 
three sets of images were co-aligned using a two-dimensional cross correlation routine. We also used 
4k $\times$ 4k-pixel, AIA \ion{Fe}{ix}~171~\AA~ and \ion{He}{ii}~304~{\AA} 
images, which have a spatial sampling of 0.6\arcsec~pixel$^{-1}$ and a cadence of up to 12~sec. 
These wavelengths correspond to temperatures of $\log(T)\!=\!5.8$ and $\log(T)\!=\!4.7$.

We investigated the chromosphere of the active region using observations from the Chromospheric Telescope 
\citep[ChroTel,][]{2008SPIE.7014E..36K,2011A&A...534A.105B, 2012A&A...537A.130B} operated next to the German Vacuum Tower 
Telescope \citep[VTT;][]{1985VA.....28..519S} at the Observatorio del Teide, Tenerife, Spain. ChroTel provides 
narrow-band filtergrams in three wavelengths, namely, \ion{Ca}{ii}~K~ 3933.7~\AA, H$\alpha$~6562.8~\AA, and \ion{He}{i}~10830.3~\AA, 
using three separate Lyot filters. The passbands of the Lyot filters at the above wavelengths are 0.3, 0.5, and 1.3~\AA, 
respectively. Around 10830~\AA, a tunable filter was employed, which delivers filtergrams at seven fixed 
wavelengths (numbered 1--7) with non-equidistant wavelength spacings. This arrangement facilitates 
the retrieval of LOS velocities. The 2k $\times$ 2k-pixel images have a spatial sampling of about 1$^{\prime\prime}$~pixel$^{-1}$ 
and exposure times of 1000, 300, and 500 ms, respectively. On July 2, ChroTel provided filtergrams in all three 
wavelengths with a cadence of 3~min from 6:54~UT to 17:06~UT. The ChroTel images were corrected for limb darkening 
and rescaled to match the HMI images. The chromospheric filtergrams were deprojected and co-aligned using the same
scheme as mentioned above.

\section{Results}
\label{result}
\subsection{General evolution of active region NOAA 11515}
\label{gen}

The leading sunspot of AR~11515 became clearly visible at the east limb on 2012 June 27. We analyse its evolution in the period July~1--4 
displayed in Figs.~\ref{figure01} and \ref{figure02} (each with co-temporal panels P1--P24). At the beginning of July 1, the region is 
located at S19 E28 and has a $\beta\gamma$ magnetic configuration \citep[following the Hale classification,][]{1938QB525.H13......}, 
consisting of a leading sunspot of positive polarity followed by a complex group of smaller spots, mostly of negative polarity.
From about this time, a group of satellite sunspots with the same polarity as the leading sunspot appears south of the latter 
(panels P1--P4). Additionally, by $\approx$\,14:00~UT, the emergence of significant new flux in front 
of the leading sunspot is obvious (P3--P5). It consists of mixed but dominantly negative polarity, thus
forming a new neutral line. This region of flux emergence and its associated neutral line develop into an arc that extends southwards to the 
emerged satellite polarity (P6-P7). A filament begins to form in the corona along the newly established neutral line around 18:30~UT, 
as seen in AIA 304~{\AA} images. 

In the early part of July 2, the leading sunspot, as well as the small group of satellite sunspots, clearly separated from their more 
complex trailing spots. During this period, the leading sunspot is seen to stretch, with penumbral filaments on either side moving towards 
each other near the middle of the spot (P8--P9). The leading edge of the spot thus approaches the newly formed neutral line, 
a situation previously found to trigger flare activity on a wide range of scales, from a large $\delta$ spot region 
producing X-class flares \citep{1991ApJ...380..282W} down to sub-arcsecond scales \citep{1998ApJ...502..493D}. Additionally, the arc-shaped 
region of dominantly negative flux emergence approaches the satellite polarities in a period of more than a day (P5--P15), producing a 
closely packed configuration that is suggestive of flux cancellation. 

At about 10:43~UT on July~2 the filament erupts and simultaneously an M5.6 flare commences. The eruption begins near the satellite 
polarities (dotted square in P9--P10) but quickly extends along the whole length of the neutral line, and results in a CME.

\begin{figure*}[!ht]
\centerline{
\includegraphics[angle=0,width = 0.95\textwidth]{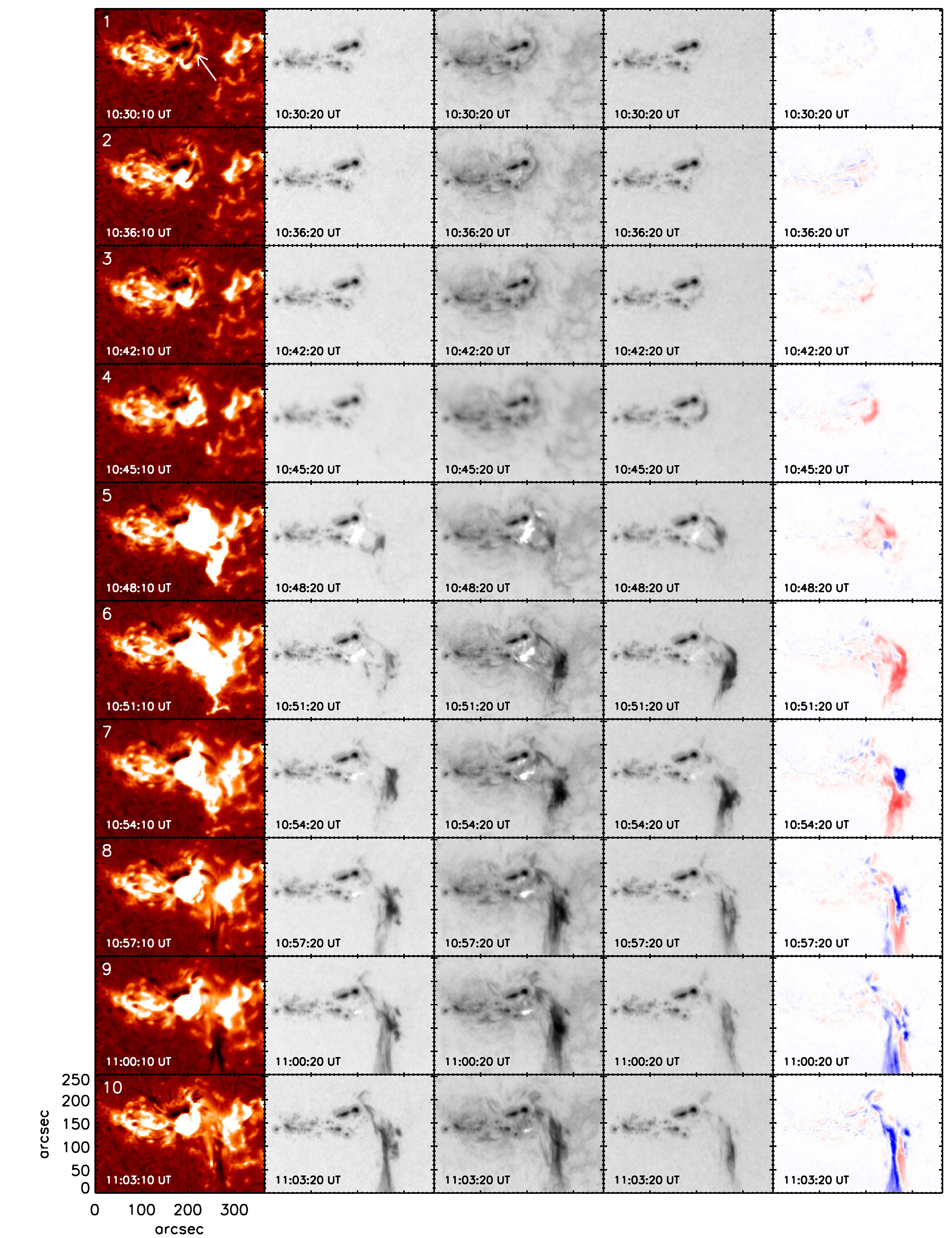}
}
\vspace{0pt}
\caption{Temporal evolution of the M5.6 flare on 2012 July 2 as seen in ChroTel observations. The left column shows 
ChroTel H$\alpha$ filtergrams, and columns two to four correspond to ChroTel filtergrams at 10828.5~\AA,  
10830.3~\AA, and 10832.1~\AA, respectively. Column~5 depicts He {\sc i} LOS velocities derived from the five innermost filtergrams. 
Blue corresponds to blue shifts and red to red shifts. The panel number is indicated in the top left corner.}
\label{figure03}
\end{figure*}

The stretching of the leading sunspot continues until it splits into two nearly equal parts (P11), and the leading half and its moat separate 
from the rest of the active region. This ``runaway'' sunspot exhibits significant proper motion and rotation after its 
separation, thus continuing to approach the arc-shaped neutral line ahead of it. Episodes of flux cancellation with minority polarity patches 
ahead of it then show up more clearly. Strong shear flows are observed near the periphery of the runaway sunspot in this phase and 
subsequently (P17--P19). These shear flows are no longer seen on July 4, and the sunspot drifts further away. The other half of the leading 
sunspot that remained within the active region merges with the satellite group of sunspots, where a strong neutral line is established early
on July 4. This newly established $\beta\gamma\delta$ configuration produces a long series of further flares, including 
20 M-class flares, several of which cause a Sun quake (S. Zharkov and J. Mart\'inez Oliveros, personal communication).

\subsection{The M5.6 flare on 2012 July 2}
\label{flare}

Although many M- and C-class flares originate in AR~11515 during its transit on the solar disc, in this 
paper we only describe and analyse the M5.6 flare and possible precursors associated with the sunspot splitting. The 
available multi-wavelength data cover the evolution of the flare and of the erupting filament comprehensively, both in the chromosphere and 
corona.

At the time of the eruption, the active region is located at S19 E8 ($\Theta$~$=$~$20.7^{\circ}$). In the 1--8~{\AA} soft 
X-ray band monitored by the Geostationary Operational Environmental Satellite \citep[GOES,][and references therein]{2011SoPh..272..319N}, the 
M5.6 flare starts near 10:43~UT and peaks at 10:52~UT. The flux decreases to the (relatively high) background level of $\sim10^{-6}$~W~m$^{-2}$ 
around 11:10~UT, although the flare is eruptive and associated with the filament eruption and subsequent CME. A 
preceding C2.9 flare started at 10:33~UT and peaked at 10:37~UT. AIA images show it to be associated with a faint, southwards propagating spray, 
but otherwise there are no indications of ejective behaviour. This suggests that the C2.9 flare is a confined event.

Figure~\ref{figure03} shows the time sequence of H$\alpha$ (column 1) and \mbox{He {\sc i}} 10830 \AA~ images from ChroTel, 
which depict the spatial evolution of the flares. Columns 2 to 4 correspond to filtergrams taken at different 
central wavelength positions, namely 10828.5, 10830.3, and 10832.1 \AA. At these wavelengths, the 
Lyot filter samples the outermost red wing of a photospheric Si~{\sc i} line, the two strongest helium lines, and a water vapour 
line on the red side of the He~{\sc i} lines, respectively. Column 5  shows the LOS velocity derived from the five 
innermost filtergrams \citep[nos. 2--6 in ][]{2011A&A...534A.105B}. The derivation of the LOS velocity has been explained in 
\citet{2011A&A...534A.105B}, wherein the weight factor for each filtergram has to be determined through 
calibration with spectroscopic data. Since there were no spectroscopic observations of the He~{\sc i} triplet on 
July~2, we compute the LOS velocity from the mean weight factors determined by \citet{2011A&A...534A.105B}. 
For this reason, and since we are dealing with a wide range of velocities from several km~s$^{-1}$ in the region of the flare to less than 
2~km~s$^{-1}$ in the quiet Sun, we refrain from making quantitative estimates of the velocities.

\begin{figure}[!h]
\centerline{
\includegraphics[angle=90,width = \columnwidth]{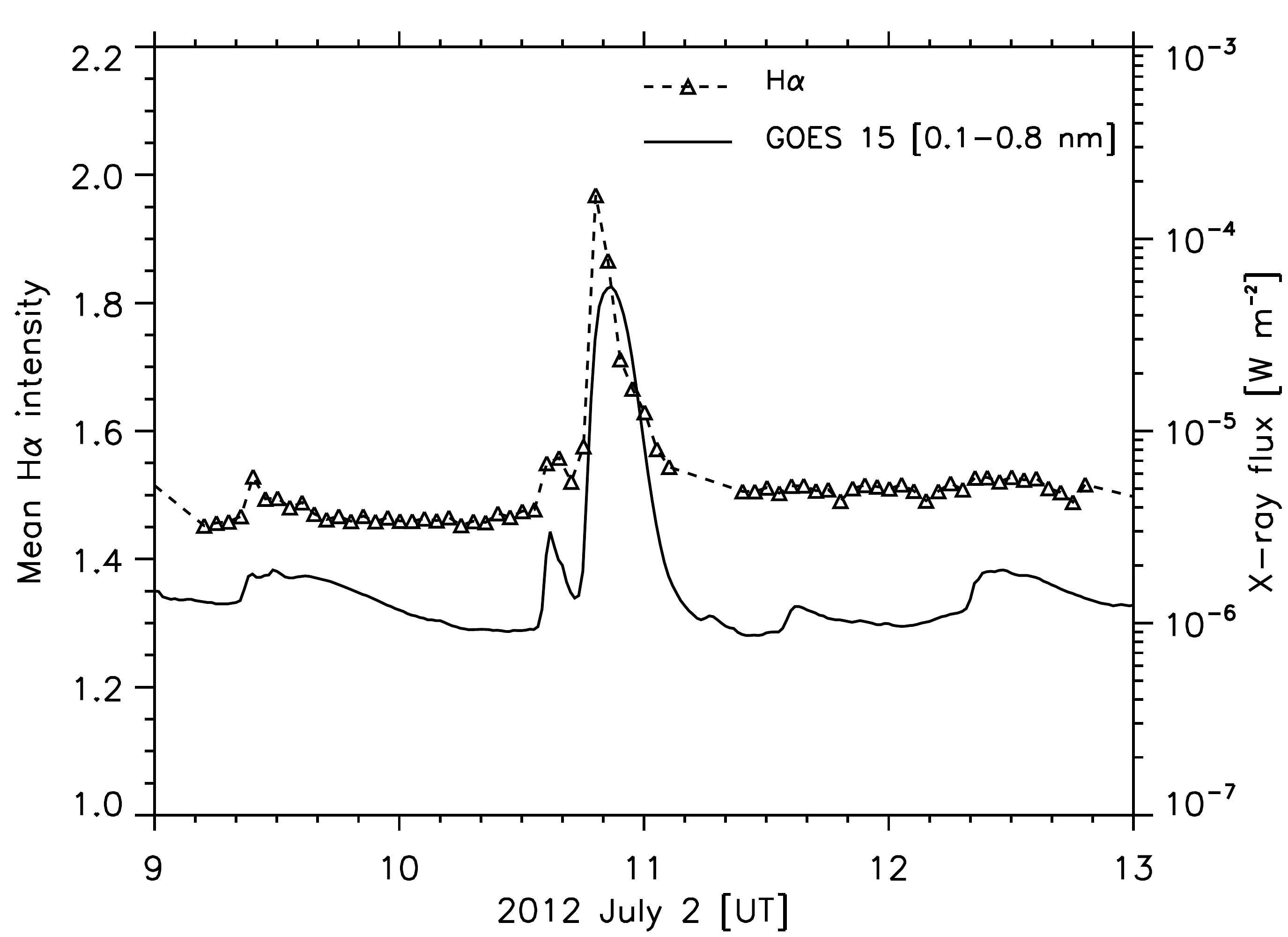}
}
\vspace{-5pt}
\caption{Temporal evolution of the mean H$\alpha$ intensity ({\em dashed} line with {\em triangle} symbols) in the region 
of the flare as obtained from ChroTel, and the GOES 15 X-ray flux in the 0.1--0.8~nm energy channel ({\em solid} line). 
A small C2.9 flare occurred at 10:37~UT, which is immediately followed by the M5.6 flare at 10:52~UT.}
\label{figure04}
\end{figure}

Panel~1 of Fig.~\ref{figure03} shows the presence of a large filament (white arrow in ChroTel H$\alpha$ image) following the arc-shaped 
neutral line in front of the leading spot and extending to the area of the satellite group of sunspots. The length of the filament is 
about 50$^{\prime\prime}$ and the width is about 20$^{\prime\prime}$. At the southern end of the filament, there is a compact 
brightening close to the satellite group of sunspots. Similar brightenings are observed near the northern end of the filament close to the 
leading sunspot, in the complex negative-polarity part of the active region and in the trailing part of AR~11514 west of the leading 
sunspot (right side of panel 1 in the ChroTel H$\alpha$ image). Six minutes later, the satellite group of sunspots is covered by H$\alpha$ 
emission that has expanded from the brightening seen earlier (panel~2). Smaller intensity enhancements are seen in ChroTel filtergrams around 
10830~\AA, which are more pronounced in the outer wing of the Si~{\sc i} line and the He~{\sc i} triplet (columns 2--3). These appear like 
small flare kernels \citep{1985AuJPh..38..875H} and are the signatures of the C2.9 flare. The H$\alpha$ emission stays at an enhanced level 
in the decay phase of this flare, while the He filtergrams are devoid of any emission at 10:42 and 10:48~UT (panels 3--4). The filament is no 
longer visible in H$\alpha$ after the peak of the C class flare (panels 4--5), possibly because of the heating and ionization of its plasma 
\citep{2012NewA...17..732Y}. We obtain an H$\alpha$ light curve of the flares from the ChroTel images by calculating the mean H$\alpha$ 
intensity within a smaller field of view enclosing the flare emission and excluding dark features such as sunspots and the filament. The 
comparison with the GOES X-ray light curve in Fig.~\ref{figure04} demonstrates that the brightening around 10:36~UT in the ChroTel images 
is the C2.9 flare, which thus has a close spatial association with the M5.6 flare, that is, it is a true precursor event. 

\begin{figure}[!h]
\centerline{
\hspace{0pt}
\includegraphics[angle=0,width = \columnwidth]{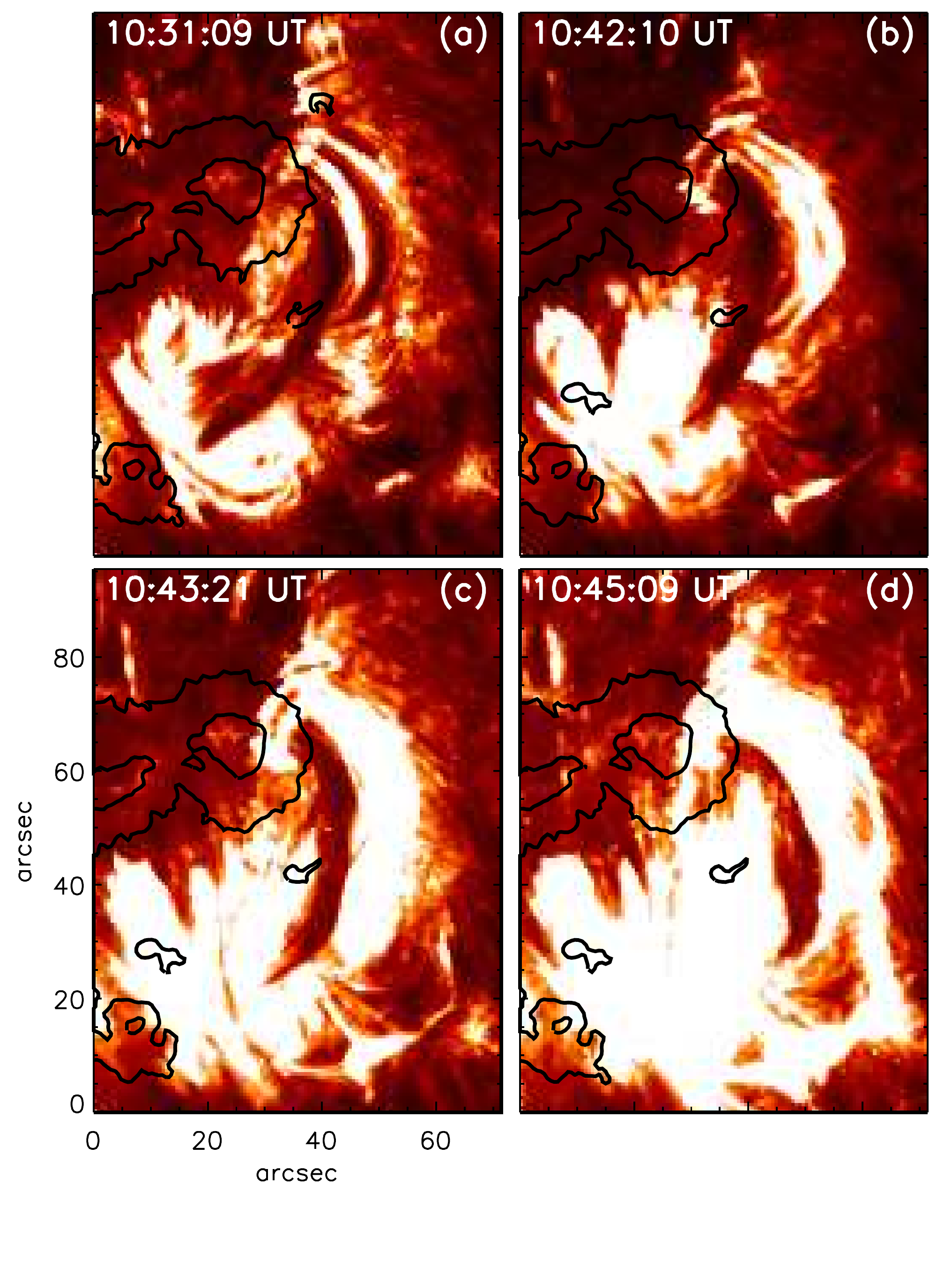}
}
\vspace{-30pt}
\caption{Sequence of AIA 304~{\AA} images showing the filament immediately before (a) and after (b--d) the onset of the 
eruption.}
\label{fig-AIA304}
\end{figure}

\begin{SCfigure*}[][!ht]
\begin{minipage}{0.74\textwidth}
\centering{
\hspace{10pt}
\includegraphics[angle=90,width = \textwidth]{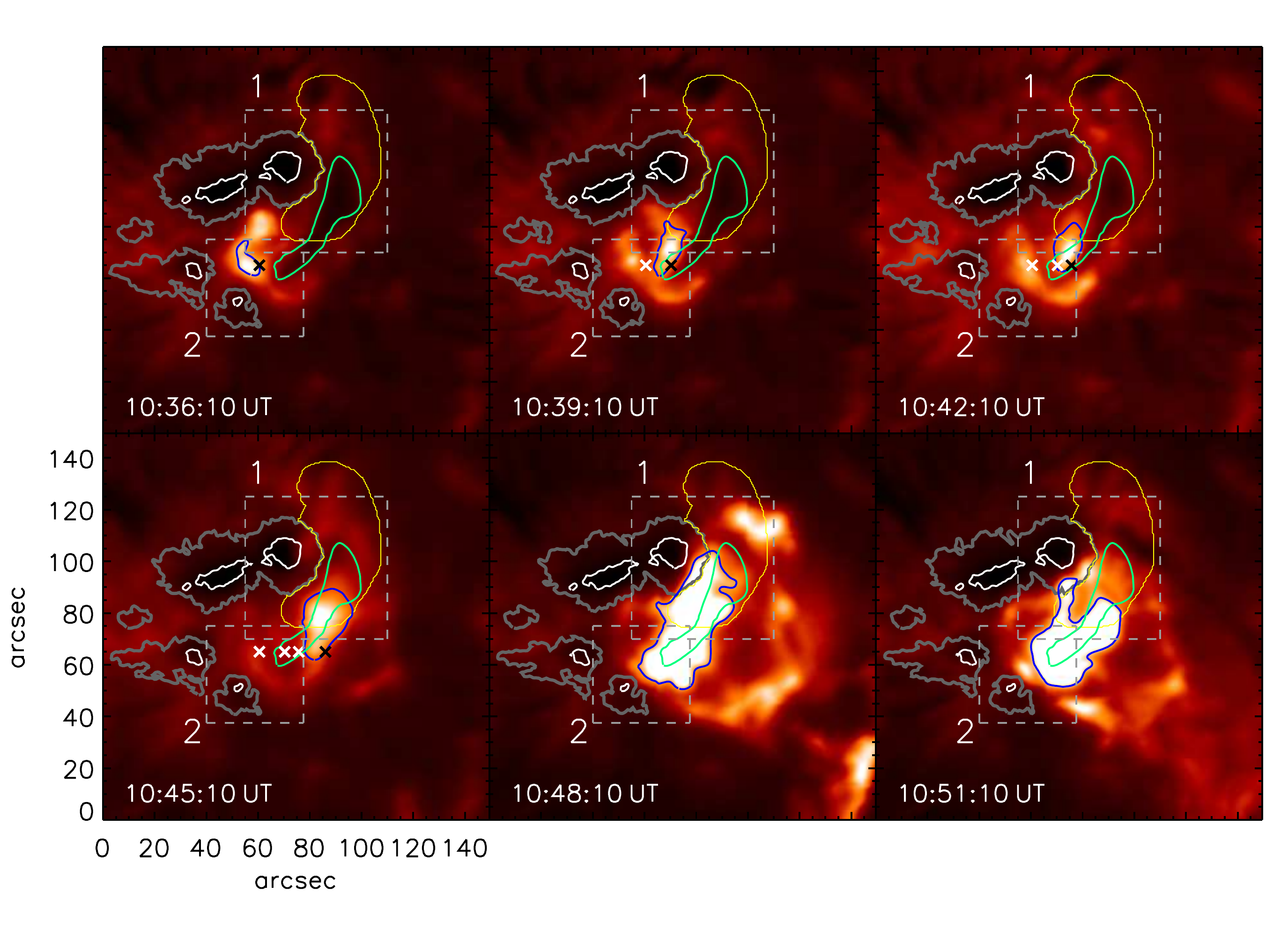}
}
\vspace{-40pt}
\caption{Evolution of the C2.9 precursor flare towards the onset of the M5.6 flare. Each ChroTel H$\alpha$ image has been individually scaled 
to its maximum intensity. The green and blue contours outline the filament and the flare brightenings, 
respectively, as observed at H$\alpha$. The white and grey contours correspond to the sunspot umbra-penumbra and penumbra-quiet Sun boundary, 
respectively. The yellow contour shows the mask used in deriving the amount of flux emerging near the neutral line, plotted 
in Figure~\ref{figure07-2}. The black and white {\em crosses} indicate the position of the leading edge of the brightness front in 
the current and previous frame, respectively. The {\em grey dashed squares} labelled 1 and 2 are small regions selected for analysing magnetic 
cancellation and/or emergence events.}
\label{figure05}
\end{minipage}
\end{SCfigure*}

The He~{\sc i} emission of the M5.6 flare in the ChroTel filtergrams commences in the impulsive flare phase 
(panel 5 of Fig.~\ref{figure03}). During the peak of the flare, a strand of the erupting filament is seen in 
H$\alpha$, while the He filtergrams capture the acceleration of the body of the filament as it erupts. In particular, 
the first changes in the location of the filament occur between the 10832.1~{\AA} images in panels~3 and 4.
The filament eruption is also seen in the He {\sc i} LOS velocities, which show the bulk of the filament material exhibiting 
high blue shifts (panels 7--10). One also finds traces of weak red shifts along narrow filament 
strands, which is consistent with the results of \citet{2012NewA...17..732Y}. The flare kernels, which are locations 
of strong emission, appear red-shifted in the He {\sc i} LOS velocity maps, possibly due to non-thermal heating of the 
plasma (panels 5--7). Traces of the filament are observed in H$\alpha$ after 10:57 UT, indicating 
subsequent cooling of the filament material (panels 8--10). 

The higher cadence of the AIA images allows us to narrow down the onset time of the filament eruption to the interval 10:42--10:43~UT, 
nearly simultaneous to the onset of the M-class flare. Since the exact onset time of the M-class flare is masked in the soft X-ray 
light curve by the preceding C-class flare, we cannot determine whether the filament eruption or the M-class flare commenced first. The filament 
remains visible in the AIA 304~{\AA} images up to about 10:46:30~UT. These show more clearly than the 10832.1{\AA} images in 
Fig.~\ref{figure03} that the eruption begins at the southern end of the filament (Fig.~\ref{fig-AIA304}).

The filament eruption evolves into a southward traveling CME with an estimated\footnote{Data available at http://cdaw.gsfc.nasa.gov/} 
start time near 11:05~UT, angular width of 125$^\circ$ and projected velocity of 313~km~s$^{-1}$. The true velocity must be considerably 
higher, since the eruption occurs not far from disc centre.

Figure~\ref{figure05} shows the evolution of the C2.9 flare towards the onset of the M5.6 flare as seen in H$\alpha$. ( 
The images at this wavelength do not show the whole extent of the filament; see Figure~\ref{fig-AIA304}).
The C2.9 flare launches an H$\alpha$ intensity front whose leading edge is tracked by crosses in the top panels of Fig.~\ref{figure05}. 
The intensity front moves at about 20 km~s$^{-1}$ towards the filament, traversing its southern part exactly in the interval 
where it starts to erupt (10:42--10:43~UT) and remaining at this position subsequently. This suggests that an interaction 
occurred between the precursor flare and the erupting filament. 

\subsection{Associated photospheric changes}
\label{flux}
In the following subsections, we analyse the photospheric changes in the active region around the time of the M5.6 flare, based on HMI 
continuum images and LOS magnetograms, and relate them to the activity in the upper atmosphere. A comprehensive picture 
includes the long-term trends in the region that persist for more than a day, as well as rapid changes shorter than an hour, 
which occur in close temporal and spatial association with the flare.

The most conspicuous feature in the active region is the splitting of the leading sunspot and the large displacement of its front fragment 
(Fig.~\ref{figure01}). The splitting begins about 20~hours before the M5.6 flare with a pronounced stretching and proceeds across the middle 
of the spot where the northern and southern penumbral segments approach each other (panels P5--P8 of Fig.~\ref{figure01}), but no light bridge 
appears. The velocity of the front part reaches about 300 m~s$^{-1}$, and two fragments of nearly equal size are produced 
about six hours after the flare. The primary component of the front fragment's motion points westwards, i.e. towards the 
arc-shaped neutral line ahead of it, where one end of the filament is anchored. The separation also has a weaker northward 
component. Since the neutral line runs roughly in the northwestern direction in this area, the splitting of the spot comprises a shear flow 
component in addition to the converging flow. This is fully analogous to previously analysed cases of flares caused by spots moving towards 
newly emerged flux in their vicinity \citep[e.g.,][]{1991ApJ...380..282W, 1998ApJ...502..493D}. (See the following sections for further 
development of the split sunspot.)

\begin{figure}[!h]
\vspace{0pt}
\centerline{
\includegraphics[angle=90,width = \columnwidth]{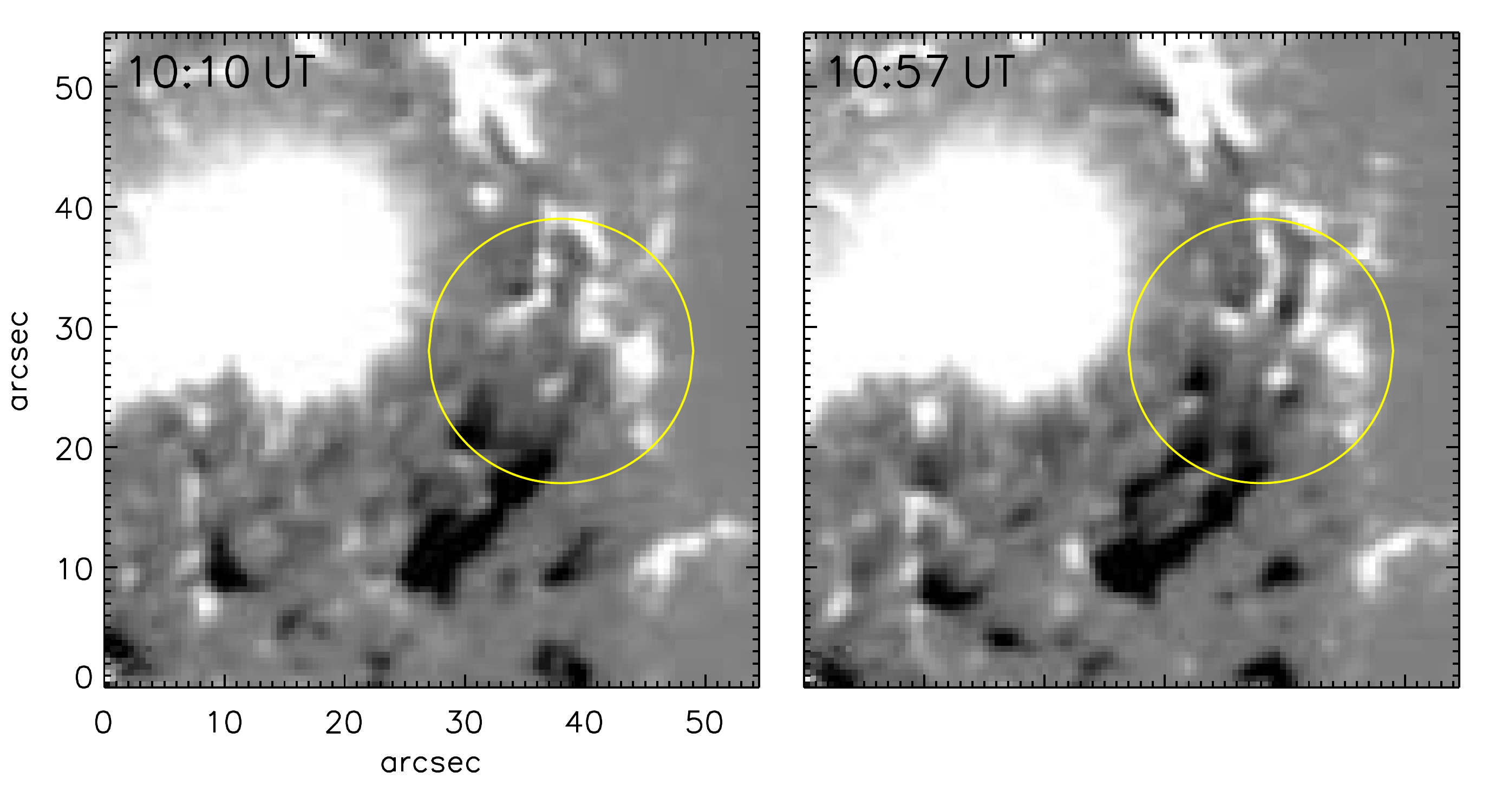}
}
\vspace{-10pt}
\caption{Events of flux emergence at the neutral line close to the leading sunspot. The field-of-view in the figure 
corresponds to the region marked by the grey dashed square labelled `1' in Fig.~\ref{figure05}. The yellow circle 
encloses patches of negative polarity that emerge close to the time of the M5.6 flare.}
\label{figure07-1}
\end{figure}
\vspace{-10pt}
\begin{figure}[!h]
\vspace{-10pt}
\centerline{
\includegraphics[angle=90,width = \columnwidth]{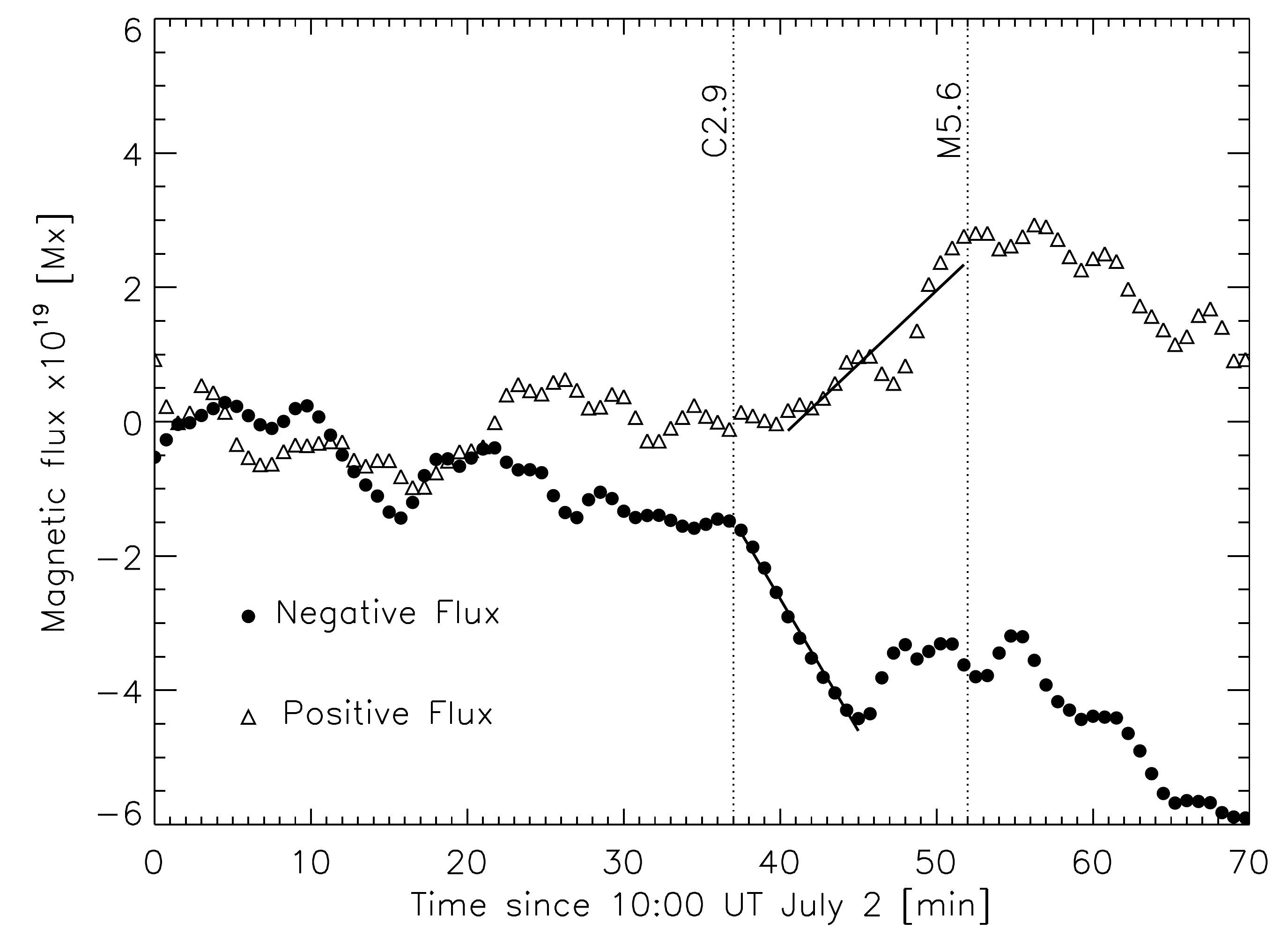}
}
\vspace{-5pt}
\caption{Temporal change of magnetic flux within in the area enclosed by the yellow line in 
Fig.~\ref{figure05}, including linear fits to the steepest parts.}
\label{figure07-2}
\end{figure}

\begin{figure}[!h]
\centerline{
\hspace{5pt}
\includegraphics[angle=0,width = 1.0\columnwidth]{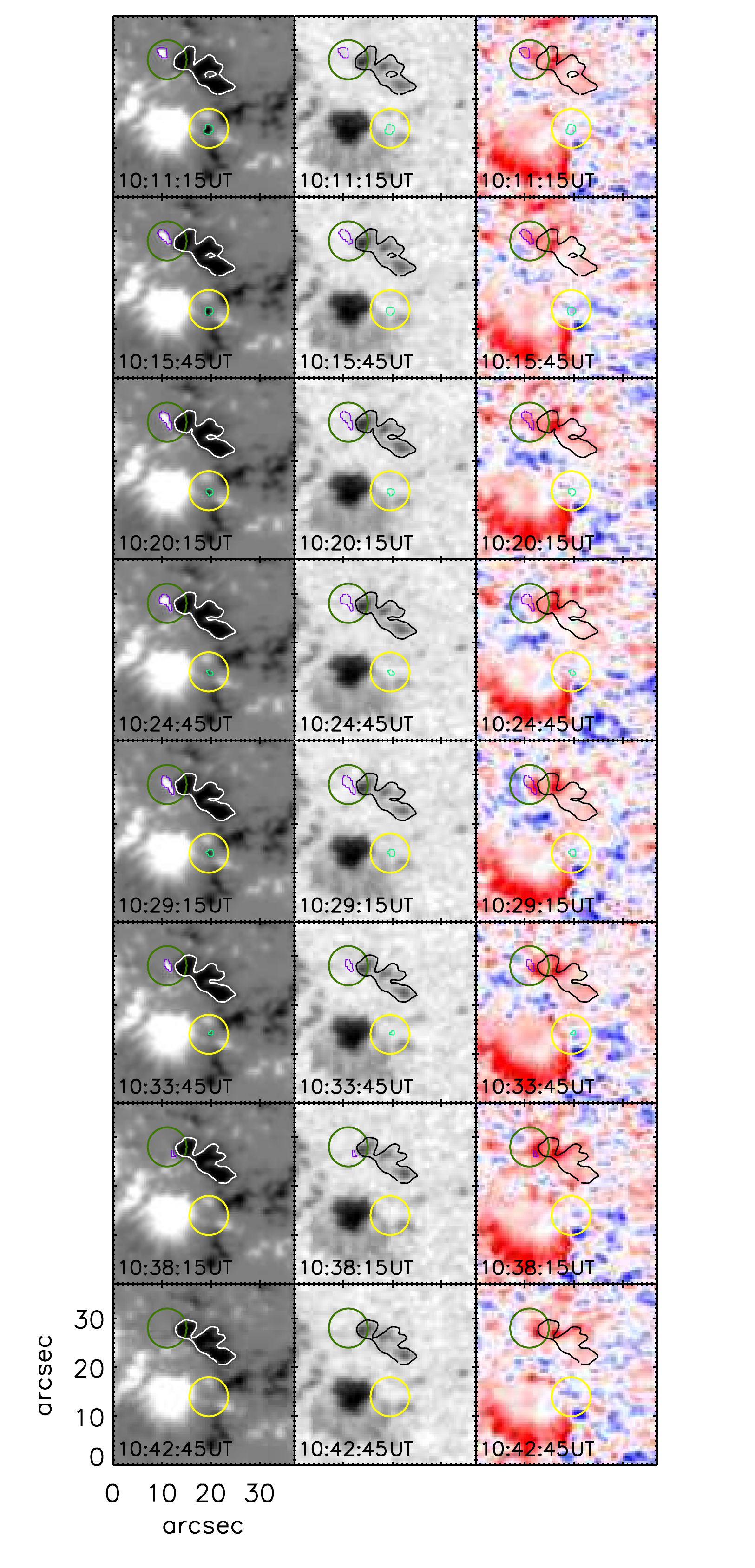}
}
\vspace{-15pt}
\caption{Photospheric magnetic flux cancellation/disappearance near the polarity inversion line. The images of the time 
sequence correspond to the LOS magnetograms (left), continuum intensity (middle), and dopplergrams (right) for the region 
marked by the grey dashed square labelled `2' in Fig.~\ref{figure05}. The green (yellow) circle encloses a positive 
(negative) magnetic patch outlined in violet (turquoise). The magnetograms, continuum intensity, and dopplergrams have 
been scaled between $\pm$1000 G, 0.1--1.05~$I_{\textrm{\tiny{QS}}}$, and $\pm$1.5 km~s$^{-1}$, respectively. The violet 
and turquoise contours enclose fields greater than 80 and 50~G, respectively, and have been enhanced by a factor of six 
for clarity. The white contour in the left column (black in columns 2--3) outlines a region of previous 
flux emergence where the flux also decreases in the time before the flares.}
\label{figure06}
\end{figure}

The evolution towards the eruption also includes many changes in the photospheric flux on smaller scales.
Figure~\ref{figure07-1} illustrates instances of small-scale flux emergence along the neutral line close 
to the leading sunspot in the box labelled `1' in Fig.~\ref{figure05}. The yellow circle outlines patches
of positive and negative polarity that emerge close to the time of the M5.6 flare. 
Figure~\ref{figure07-2} shows the temporal change of the magnetic flux near that neutral line. All flux 
detected within the yellow contour in box `1' in Fig.~\ref{figure05} is plotted, after subtracting the fluxes at 10:00~UT. 
Most of the yellow contour in Fig.\ref{figure05} is selected by eye to include all significant contributions and fixed throughout the time series, 
but the section at the periphery of the splitting spot is determined automatically at the outer penumbra boundary in the 
continuum image at the time of each measurement. The plot in Fig.\ref{figure07-2} shows a close temporal association of flux emergence 
with the triggering of the M flare. While the peak emergence rates of the two polarities differ moderately ($3.7\times10^{16}$~Mx~s$^{-1}$ 
and $-6.6\times10^{16}$~Mx~s$^{-1}$ from the linear fits in the figure), the total changes preceding the M5.6 flare 
are about the same, nearly $\pm3\times10^{19}$~Mx. 

Complex changes likewise occur near the part of the neutral line north of the satellite spots, i.e. in the area of the 
precursor flare and the initial filament motion. Both the satellite spots (positive flux) and the adjacent southern end of the 
arc-shaped flux emergence region (negative flux) are active sites of flux emergence prior to and throughout the flares 
(Sect.~\ref{gen} and Fig.~\ref{figure02}). Additionally, these areas of opposite polarity approach each other 
in a period of about 40~hr that includes the flares (panels P5--P15 in Fig.~\ref{figure02}). 
The latter evolution is highly suggestive of flux cancellation, but very difficult to quantify, since the presumed 
cancellation is superimposed on flux emergence. However, several small-scale flux cancellation episodes embedded in this 
evolution can be clearly discerned, and we present two of them in the following. The flux cancelled in these episodes is only 
a small fraction of the whole flux that disappears at this part of the neutral line in the 40~hr interval 
(compare Fig.~\ref{figure02}).

Figure~\ref{figure06} shows a magnified view of a small region marked by the {\em grey dashed square} and labelled 
`2' in Fig.~\ref{figure05}. Two sites of flux cancellation/disappearance were identified. 
The two small magnetic patches have been enhanced in flux by a factor of six in this plot
for ease of identification. Both patches appear close to 10:00~UT. In the first event, a positive flux patch 
marked by the violet contour lies adjacent to the newly emerged flux region of opposite polarity, 
which has been indicated by the white contour. This patch appears to move closer to the latter with a velocity of about 
630 m~s$^{-1}$, traversing almost the full diameter of the green circle shown in Fig.~\ref{figure06}. 
This velocity is consistent with the range of 300--800~m~s$^{-1}$ estimated by \citet{2000SoPh..197...75M}. 
The magnetic flux increases slowly from $2.3\times10^{18}$ to $3.3\times10^{18}$~Mx in a time span of 30~min starting at 10:00~UT. 
However, this is followed by a rapid decrease in flux from 10:30~UT as is evident in Fig.~\ref{figure06}, with the patch 
completely disappearing at 10:39~UT. 

The other event identified in the sequence is observed close to the edge of the satellite sunspot's penumbra. 
There, a negative flux patch, shown by the turquoise contour, appears to be an island surrounded mostly by positive flux. 
Different from the positive flux patch described above, this negative polarity patch appears to be stationary throughout its lifetime. The 
initial flux was estimated to be $-1.9\times$10$^{18}$~Mx, which steadily approached zero at 10:34 UT when it was 
no longer visible. 

\begin{figure}[!h]
\centerline{
\includegraphics[angle=90,width = \columnwidth]{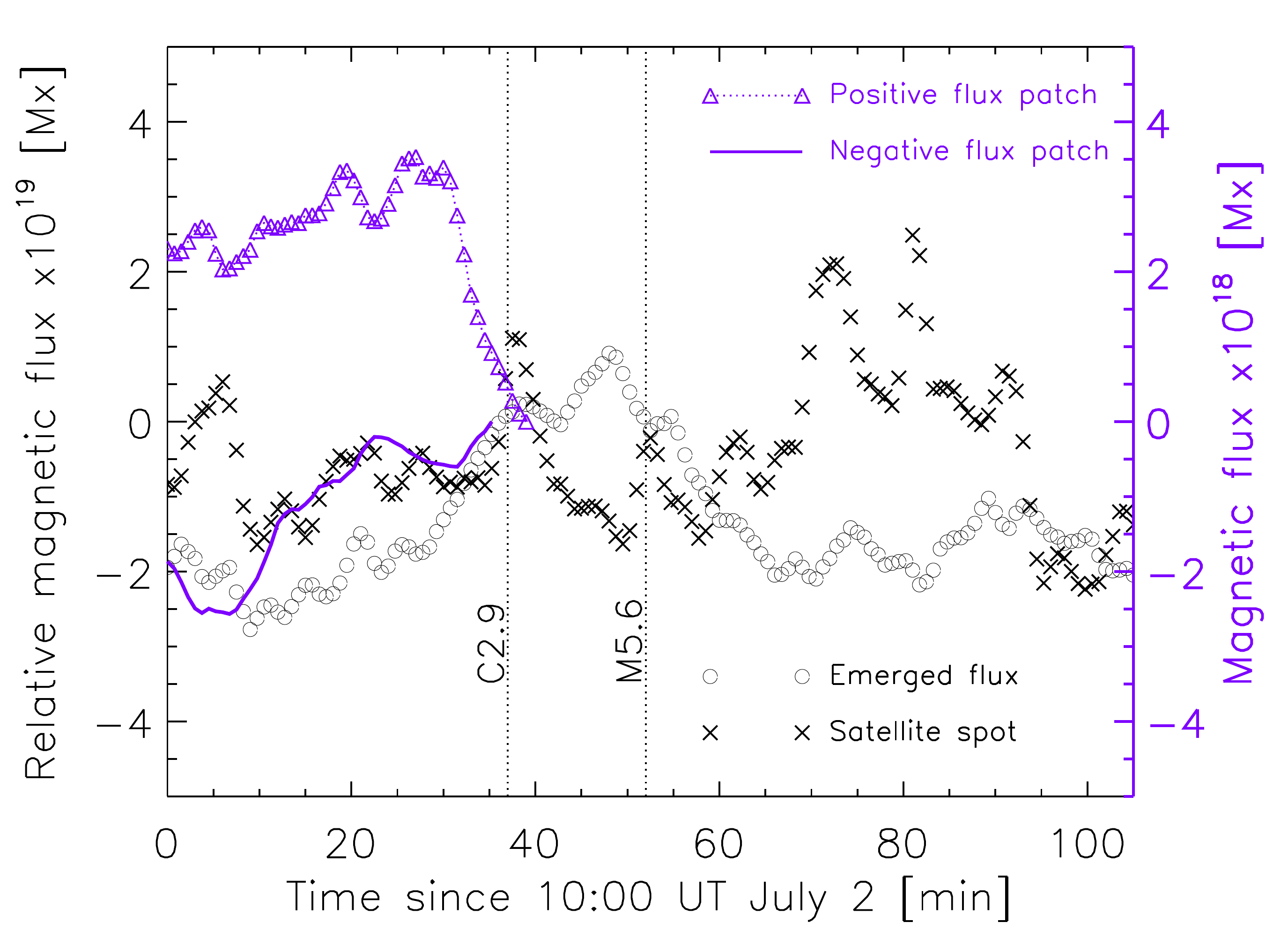}
}
\vspace{-5pt}
\caption{Temporal evolution of magnetic flux. The magnetic flux of the positive and negative polarity patches identified in Fig.~\ref{figure06} 
is shown as {\em dotted} lines with {\em violet triangles} and {\em thick violet} lines, respectively. Their scaling is shown on the violet 
$y$-axis to the right. The relative magnetic flux of the previously emerged negative-polarity region 
near the new satellite sunspot is shown as {\em open circles}. {\em Black crosses} show the positive flux in the 
satellite spot in the area shown in Fig.~\ref{figure05}. The last two quantities are referenced to their values at 10:52~UT.}
\label{figure07}
\end{figure}

Figure~\ref{figure07} shows the temporal evolution of magnetic flux in the two small patches described above. The positive and negative 
fluxes are depicted, and their values correspond to the violet $y$-axis on the right. As described above, there is an abrupt reduction 
of flux in the positive polarity patch close to the time of the C2.9 flare. 

The flux of the previously emerged negative-flux region outlined by white contours in the first column of Fig.~\ref{figure06} is shown 
in Fig.~\ref{figure07}. Here the relative magnetic flux was calculated with respect to its value at 10:52 UT at 
the peak of the M5.6 flare ($F_{\textrm{\tiny{n}}}(t) - F_{\textrm{\tiny{n,\,ref}}}$ , where ``ref'' indicates the peak time of the flare and 
$F_{\textrm{\tiny{n,ref}}} = -2.0\times10^{20}$~Mx). Starting at 10:30 UT, the relative flux steeply decreases until 10:49~UT, i.e. until shortly 
after the M5.6 flare onset. After this point, the flux first increases again to a value of $-2.1\times$10$^{19}$~Mx at 11:06~UT, but in the 
long run much of it cancels out (Fig.~\ref{figure02}, panels~P12--P15). To compare this reduction in negative flux prior to the M5.6 flare, we 
also determined the positive flux in the satellite spot (within the field-of-view displayed in Fig.~\ref{figure05}). 
In this case, the relative flux is calculated as before, namely, $F_{\textrm{\tiny{p}}}(t) - F_{\textrm{\tiny{p,\,ref}}}$,
where $F_{\textrm{\tiny{p,\,ref}}} = 5.1\times10^{20}$~Mx. The positive flux exhibits fluctuations around a 
nearly stationary average value rather than a systematic change with time. The fluctuations include minor peaks at the times of the C2.9 and 
M5.6 flares, but the peak values are not correlated with the magnitudes of the flares. Similar to the negative flux patch next to it, the flux 
in the satellite spot is determined by both continuing emergence and cancellation, resulting in the dominance of fluctuations on the time scale 
of the plot. On a longer time scale, the satellite spot is dominated by continuing emergence (Fig.~\ref{figure02}).

\subsection{Properties of the runaway sunspot}
\label{run}
\begin{figure}[!h]
\centerline{
\includegraphics[angle=90,width = 0.95\columnwidth]{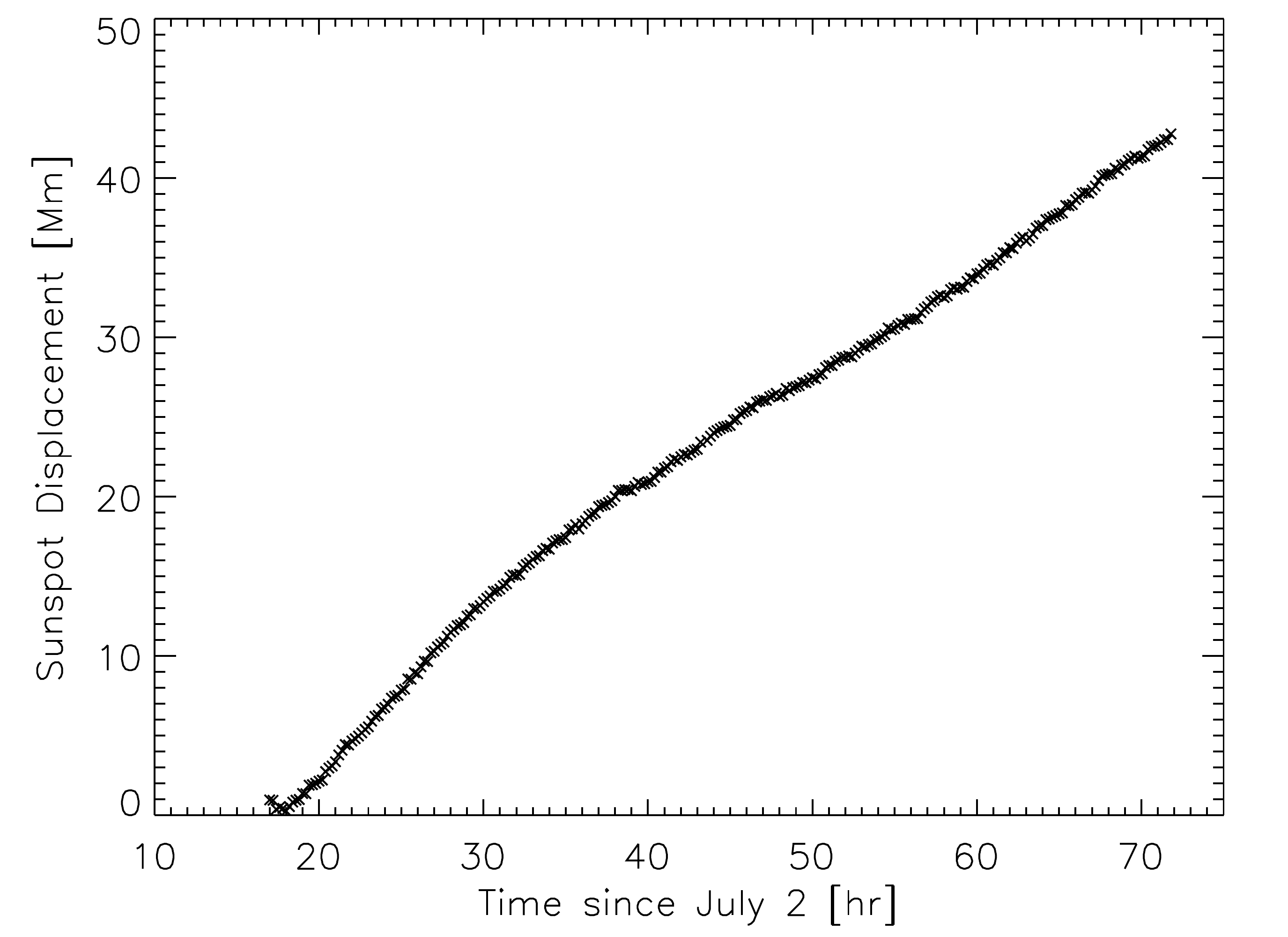}
}
\centerline{
\includegraphics[angle=90,width = 0.95\columnwidth]{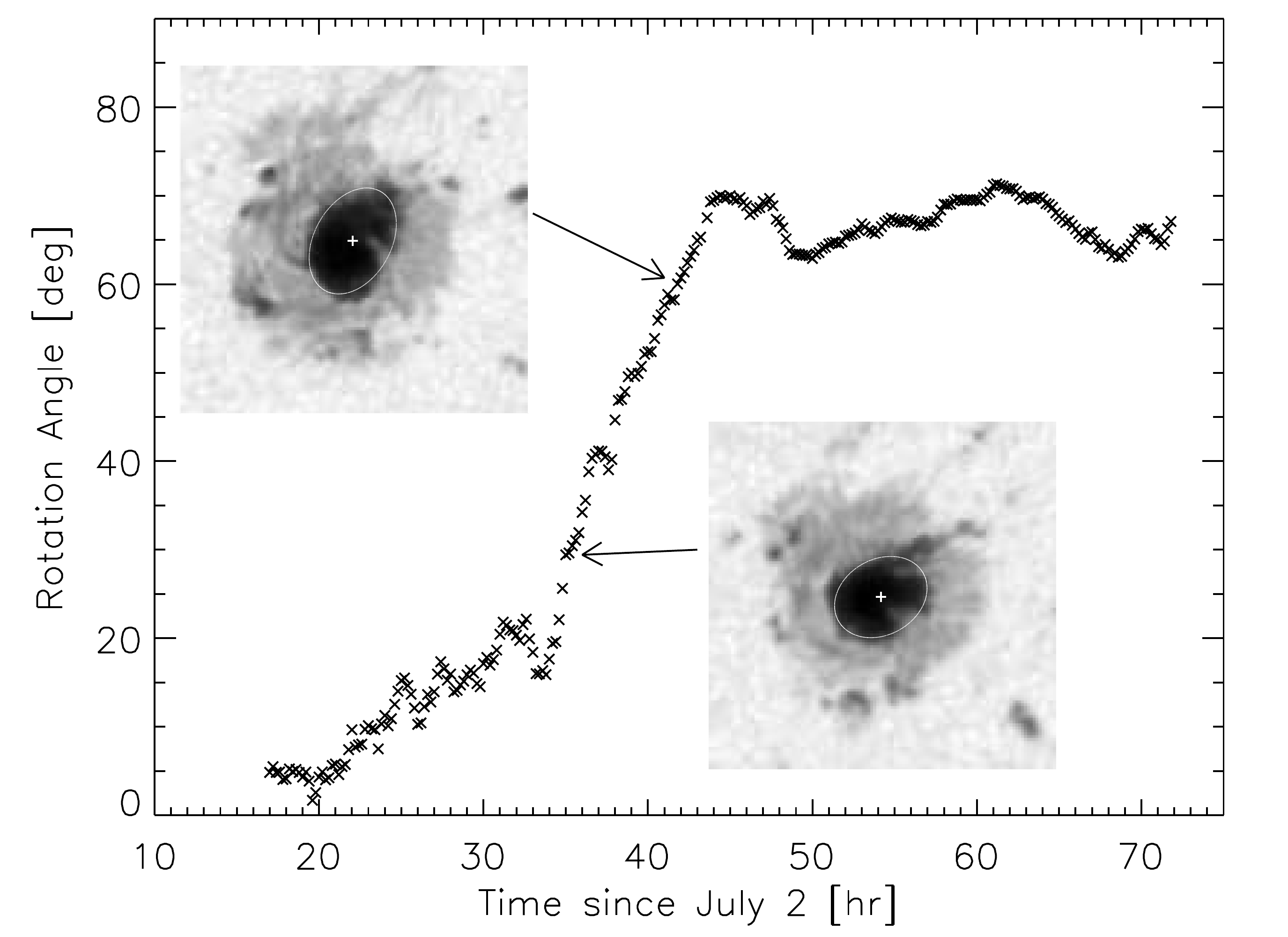}
}
\vspace{-5pt}
\caption{Displacement and rotation of the runaway sunspot are shown in the top and bottom panels, respectively, after 
separation from its host area around 16:00 UT on July~2. In the bottom panel the continuum images at two different instances along 
the steep part of the plot are shown. The thin white lines mark the umbra-penumbra boundary retrieved from fitting an 
ellipse to the observed contour and the {\em white pluses} represent the centre positions of the ellipses.}
\label{figure09}
\end{figure}

We follow the runaway part of the leading sunspot after the splitting by tracking its centroid position. Furthermore, we 
fit an ellipse to the umbra-penumbra boundary and retrieve its centre, major and minor axes, and its rotation. 
Using the above parameters, one finds that the runaway sunspot drifts about 43~Mm from its initial position
in a duration of 60 hr after the flare, moving with a nearly constant speed of about 210 m~s$^{-1}$ (top panel of Fig.~\ref{figure09}). 
The sunspot also exhibits significant rotation. The rotation angle is measured with
respect to solar west (positive $x$-axis) and increases counter-clockwise. The bottom panel of Fig.~\ref{figure09} 
shows that the sunspot rotates by nearly 70$^\circ$ in the course of 60 hr with a maximum angular speed of 
4.9$^\circ$~hr$^{-1}$ on July 3, which was estimated from a linear fit to the steepest part of the plot. 
This equals the highest rotation rate of a sunspot found so far \citep{2008SoPh..247...39Z}. Following 
\citet{2003SoPh..216...79B}, the linear speed in the outer penumbra, 20$^{\prime\prime}$ from the centre of the 
sunspot, is about 360~m~s$^{-1}$. By comparison, the other half of the leading sunspot appears to be fairly stationary 
and does not exhibit any discernible rotation.

Overall, we find a complex pattern consisting of flux increase (emergence) at the northern end of the filament and simultaneous increase and 
decrease in the flux (including cancellation) in several patches near the southern end of the filament. Changes of the order of 
$\pm(2\mbox{--}4)\times10^{19}$~Mx occur within 30~min prior to the eruption. These are embedded in longer trends of flux change associated 
with flows converging towards the neutral line both in the bow-shaped arc ahead of the leading sunspot and in the part adjacent to the satellite 
polarity.

\begin{figure*}[!ht]
\centerline{
\includegraphics[angle=90,width = \textwidth]{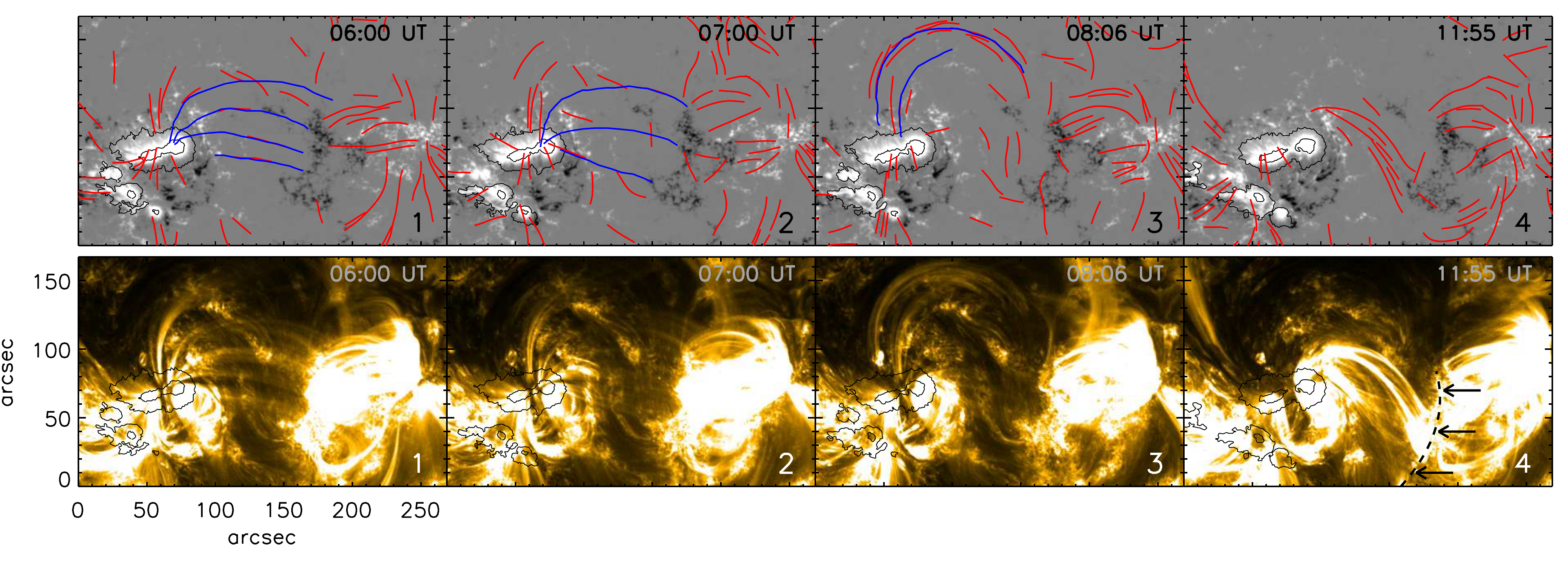}
}
\vspace{-10pt}
\caption{Evolution of coronal loops in AR~11515. The top and bottom panels show HMI LOS 
magnetograms and AIA 171~\AA~ images, respectively. The loops identified in the AIA images using the automated identification routine 
have been overlaid on the magnetograms and are shown in red. The blue loops represent those that were traced manually.
The black contours correspond to the continuum intensity and outline the sunspots in the active region. The dashed line and 
arrows in panel 4 correspond to a distinct ribbon-like brightening in the course of the M5.6 flare.}
\label{figure08}
\end{figure*}

Using coronal loops to trace the basic magnetic structure of the region, we find that the two fragments of the splitting spot possess 
completely different connections to the ambient flux. Figure~\ref{figure08} and the accompanying animation\footnote{Available as online material} 
display AIA 171~\AA~images before and after the flares. The loops in the AIA images were identified 
using the procedure described by \citet{2010SoPh..262..399A}. The routine, originally developed for Transition Region and Coronal Explorer 
\citep[TRACE,][]{1999SoPh..187..229H} images works equally well for the AIA 171~\AA~ images with appropriate adjustments of a few input 
parameters\footnote{The code is available at http://www.lmsal.com/$\sim$aschwand/software/ and is also included in the Solar Soft analysis 
suite under \$SSW/packages/mjastereo/idl/}. 
The loops in the AIA images, identified automatically and by hand, have been overlaid on the photospheric LOS magnetograms. 
These images show that all coronal loops rooted in the 
rear part of the leading sunspot connect to the following negative polarity of AR~11515, while all loops rooted in the front part, although 
diffuse and fainter, clearly connect to the west side (blue loops in panels 1 and 2 of Fig.~\ref{figure08}) where the diffuse 
trailing negative polarity of AR~11514 and enhanced negative network flux south of AR~11514 are located. The existence of the latter 
connections is confirmed  by the development of a distinct ribbon-like brightening in this area in the course of the M5.6 flare (panel~4). A 
new set of bright loops and enhanced diffuse brightness, which are obvious after the rising filament has 
largely faded (panel~4), extend from the ribbon to the leading sunspot and the arc-shaped neutral line, demonstrating that these structures 
are magnetically connected. The lack of loop connectivity between the rear half of the leading sunspot and the negative flux in and around 
AR~11514 is furthermore evident from the blue loops in panel 3 of the figure. The panel shows that the longer blue loop connects 
to a positive flux patch north of the leading sunspot while the shorter blue loop points to the middle of the splitting sunspot (not 
to its rear half).

The AIA images in the course of the flare also show that the great majority of the loops connecting the rear part of the splitting sunspot with 
the trailing negative flux of the active region do not experience any change (apart from a brief oscillation of the southernmost ones); see the 
animation accompanying Fig.~\ref{figure08}. Only few new bright loops become visible, presumably located in the interface to the overlying 
flux. This substantiates the conclusion that the two parts of the splitting sunspot have different magnetic connections.

\subsection{Shear flows around the runaway spot}
\label{shear}
As discussed in Section~\ref{flux}, the splitting of the sunspot involves a shear flow component relative to the neutral 
line under the northern end of the filament in addition to the approach to the neutral line. No shear flows are seen \emph{within} the 
splitting spot before and around the time of the flare. This changes on July~3, when the runaway sunspot approaches the opposite polarity 
patches closely.  We infer the shear flows using local correlation tracking 
\citep[LCT,][]{1986ApOpt..25..392N,1988ApJ...333..427N,2004ApJ...610.1148W,2008ASPC..383..373F}. 
After several trials in selecting the parameters of the LCT technique, an apodizing window having a width of 
4$^{\prime\prime}$ and a time difference of 12 min between the images were chosen. Three successive velocity images were averaged to reduce 
the noise in the measurements. 

\begin{figure}[!h]
\centerline{
\includegraphics[angle=90,width = \columnwidth]{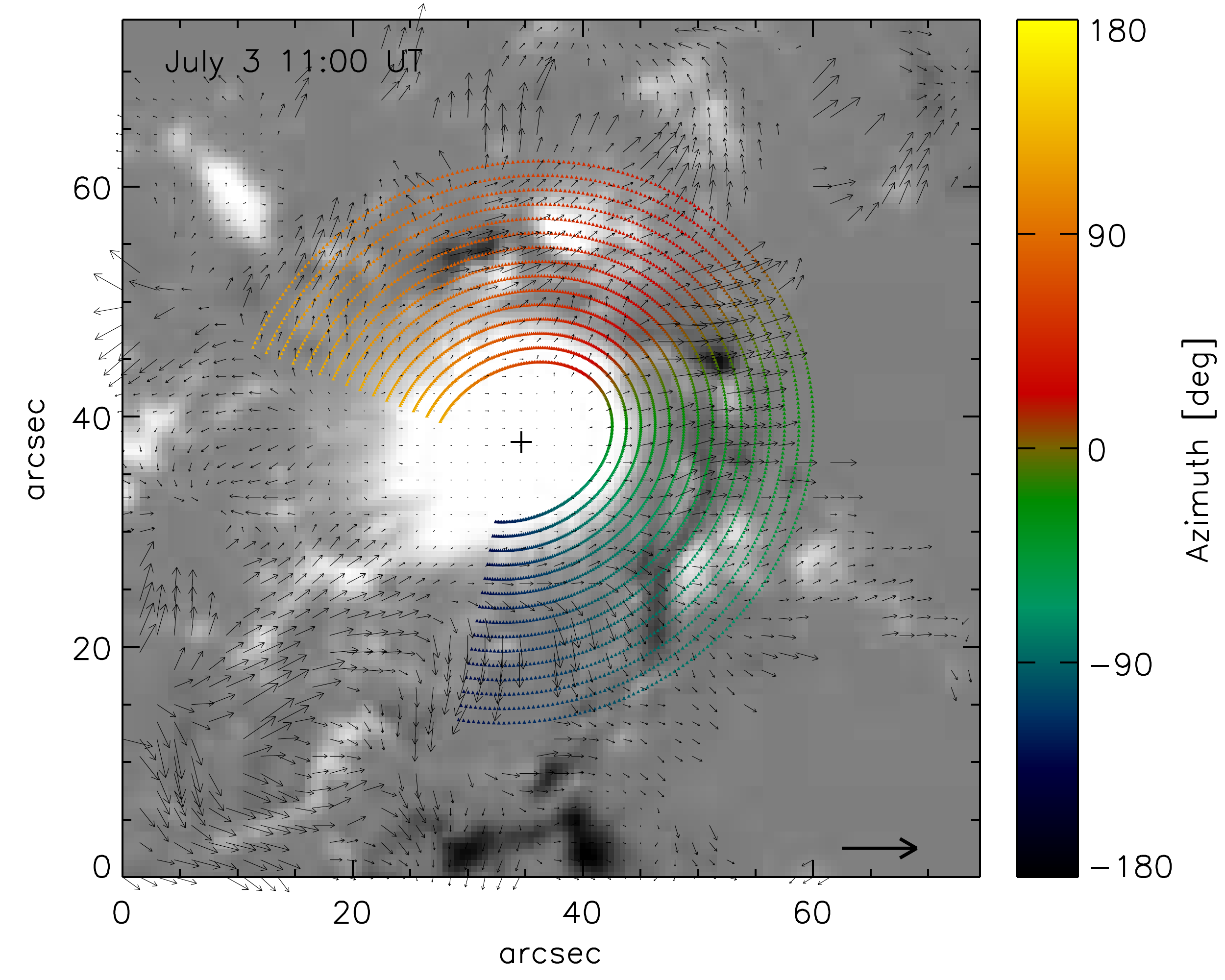}
}
\vspace{-5pt}
\caption{Horizontal flow field in the runaway sunspot (relative to its centre). The arrows indicate the flow vectors derived from LCT, which 
have been overlaid on the LOS magnetogram. The {\em thick black} horizontal arrow at the bottom right corner corresponds 
to 1~km~s$^{-1}$. The equidistant elliptical contours span an azimuth range of $-130^\circ$ to 130$^\circ$, which is 
scaled according to the vertical colour bar. A value of 0$^\circ$ azimuth coincides with the positive semi-major axis 
and increases in counter-clockwise direction. The {\em plus} corresponds to the centre of the sunspot.}
\label{figure10}
\end{figure}

Using the parameters of the ellipse that traces the umbral boundary (Section~\ref{run}), we construct 15 equally spaced 
azimuthal contours starting from the umbra-penumbra boundary and progressively displaced radially outward. Zero azimuth $\phi$ on the 
elliptical contour starts at the positive major axis and increases in the counter-clockwise direction. This is illustrated in 
Fig.~\ref{figure10}, where the equidistant contours have been overlaid on the LOS magnetogram and the horizontal 
flow vectors. The value of the azimuth is given in the adjacent colour bar. As the runaway sunspot rotates, the 
contours rotate along with it, since the tilt of the ellipse is known. We restrict our analysis to 
$0^{\circ}\leq\phi\leq130^{\circ}$ and $230^{\circ}\leq\phi\leq360^{\circ}$ and calculate the shear angle $\phi_s$ 
on the contours as the one made between the flow vector and the normal to the contour. The value of $\phi_s$ can vary 
from 0--180$^\circ$, where $0^{\circ}\leq\phi_s\leq90^{\circ}$ and $90^{\circ}<\phi_s\leq180^{\circ}$ correspond 
to outward and inward flows, respectively. 

\begin{figure}[!h]
\centerline{
\includegraphics[angle=90,width = \columnwidth]{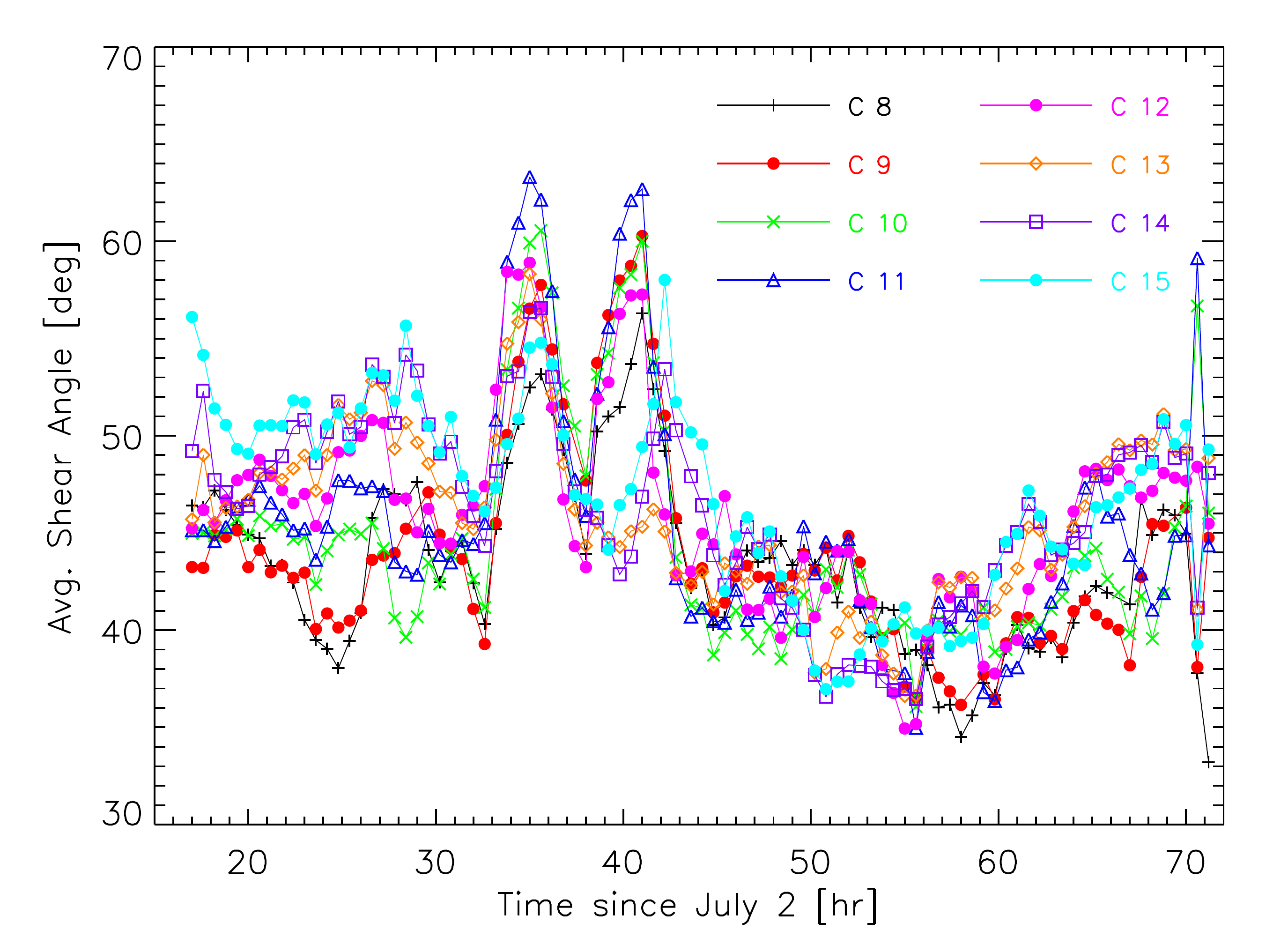}         
}
\centerline{
\includegraphics[angle=90,width = \columnwidth]{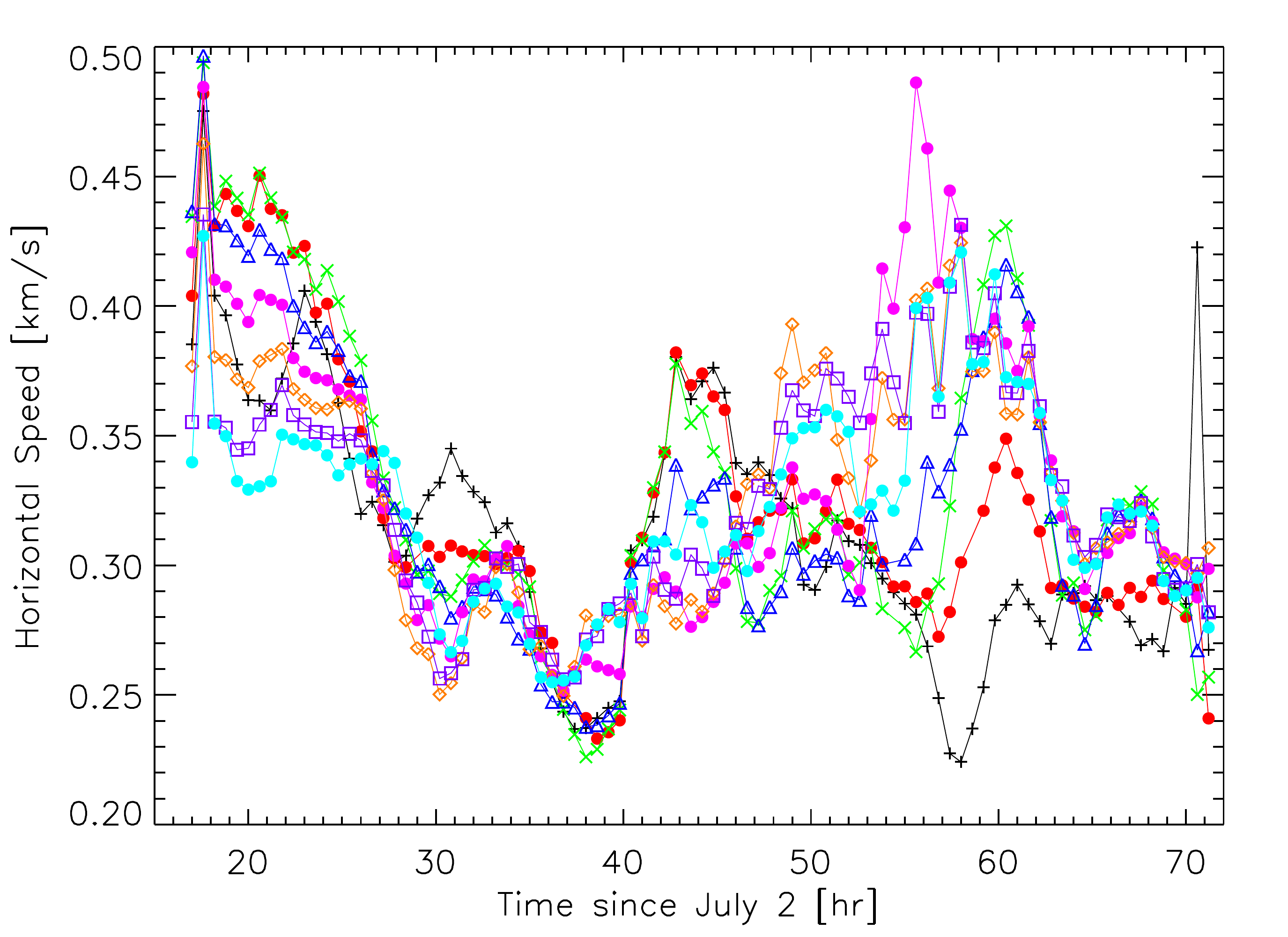}      
}
\centerline{
\includegraphics[angle=90,width = \columnwidth]{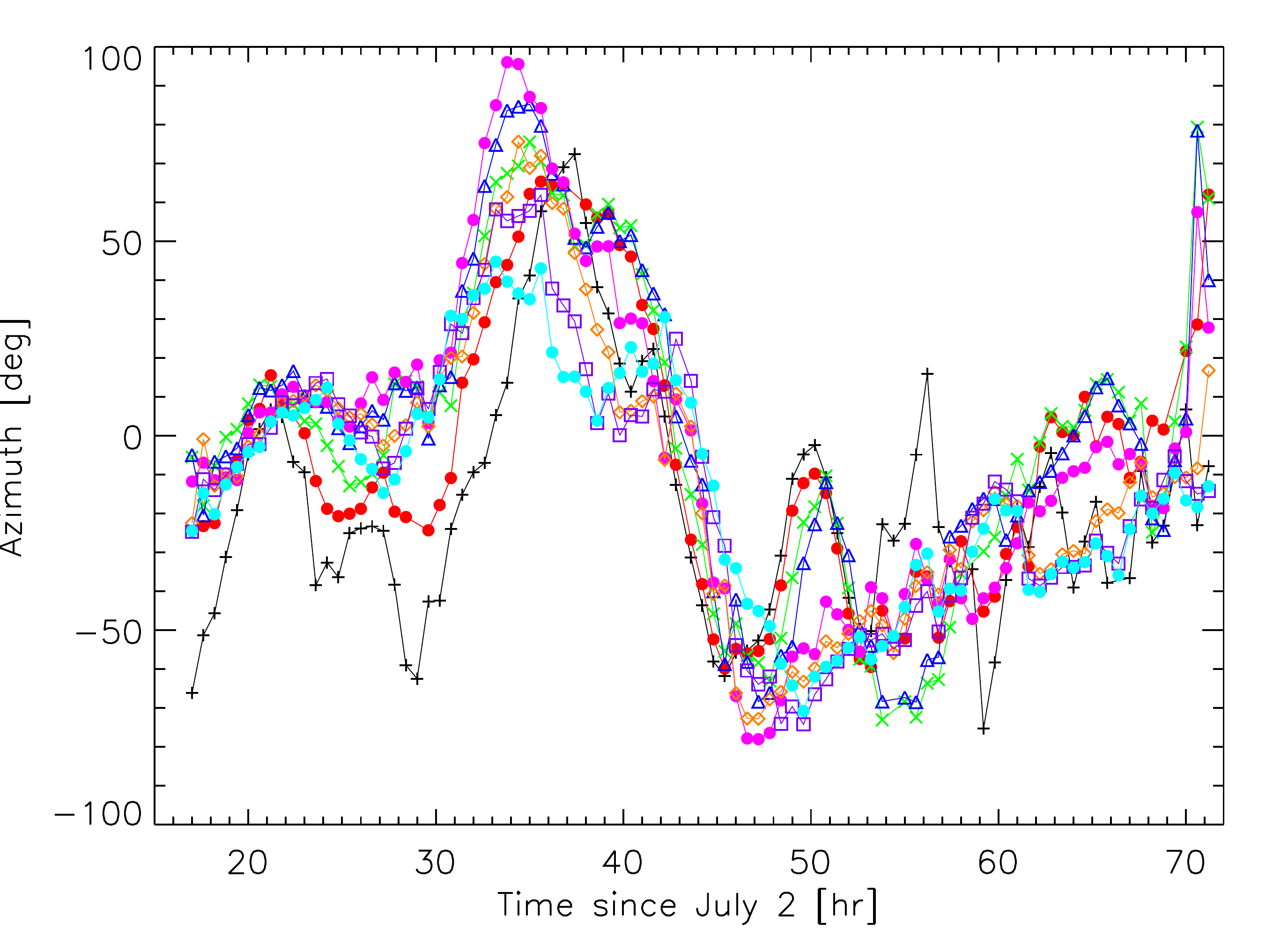}           
}
\vspace{-5pt}
\caption{Temporal evolution of shear flow around the runaway sunspot. Top to bottom: shear angle, horizontal flow speed, 
and azimuth. The azimuthally averaged values for the different equidistant radial contours are depicted in colour, 
and the legend is provided in the top panel.}
\label{figure11}
\end{figure}

The top panel of Fig.~\ref{figure11} shows the temporal evolution of 
the shear angle, averaged over the azimuth, for the eight outermost contours. The corresponding mean 
horizontal velocity is plotted in the middle panel. A representative azimuth position of strong shear flows, 
given in the bottom panel, is obtained by averaging the azimuth values of all pixels on the contour where 
$30^{\circ}\leq\phi_s\leq90^{\circ}$. 

Although one does not observe a strong time dependence in the shear flow, large shear flows ($\phi_s >60^{\circ}$) are 
predominantly seen on July 3. The mean shear angle in the outer penumbra of the runaway sunspot is about 
45$^\circ$, while the horizontal speed ranges from to 220--500 m~s$^{-1}$ with mean values of about 325~m~s$^{-1}$. In 
addition, large shear angles are not associated with strong flow speeds, which is evident for two instances on July~3 
where the speeds are about 300 m~s$^{-1}$. The bottom panel of Fig.~\ref{figure09} indicates that the shear flows span 
a large section of the sunspot's circumference during the course of two days and are not confined to a specific range 
of azimuths. There is, however, an antisymmetric trend around 0$^\circ$ azimuth wherein the shear flows change azimuth 
from 100$^\circ$ to $-80^\circ$ close to the end of July 3. The nature of the shear flows is also reflected, however 
weakly, in the relative azimuthal lag between the inner and outermost contours. This can be seen towards the 
end of July 2 and in the early parts of July 3 and July 4.

\section{Discussion}
\label{discuss}
\subsection{Triggering of the flare}
\label{triggering}
The considered event exemplifies the possible complexity in the evolution towards solar eruptions. It reveals all three 
basic photospheric driving processes to be at work, includes a precursor flare, and additionally involves a sunspot that splits in an unusual 
manner. Given this complexity, it is not clear whether a dominant causal relationship between one of the driving processes 
and the eruption can be isolated, and we refrain from stretching the interpretation of the data that far. Rather, we discuss how the various 
processes may contribute to the destabilization of the coronal magnetic field observed as the filament eruption, M-class flare, and CME.

The two major photospheric changes prior to the eruption are the splitting of the active region's leading 
sunspot and the emergence of flux that forms a group of satellite sunspots and a new neutral line ahead of the leading 
sunspot and the satellite spots. The erupting filament forms at this neutral line. 

While the splitting of the leading spot and subsequent separation of its front half represent the major change of the region 
in white light, they may have only a secondary effect on its large-scale magnetic topology. This is suggested by the fact that 
the two fragments have completely different connections to the ambient photospheric flux sources, indicating that the 
separation is not favorable for a topology change (the flux of the front half of the leading sunspot connects 
westward while the flux of its rear half connects eastward). Thus, it is possible (although not proven by our analysis) that the splitting 
of the spot neither changes the large-scale topology nor enforces the approach and reconnection of flux within the central 
part of the active region. (Verifying this conjecture would require sophisticated modelling of the evolving coronal field 
comprising two active regions, which is beyond the scope of the present paper). 

However, the splitting causes the approach of the separating front half to the new arc-shaped neutral line formed by emerging flux ahead of 
it. Seen from the neutral line, this is equivalent to a long-lasting inflow. Such inflow is crucial for the formation and destabilization of 
a coronal flux rope above the neutral line by magnetic reconnection if the field above the neutral line is sheared 
\citep{1989ApJ...343..971V,1992SoPh..138..257I,2011ApJ...742L..27A}. The shearing is realized by the northward component of 
motion of the front half. In decaying active regions, inflows towards the neutral line are always associated with flux 
cancellation, but here flux emergence is dominant and prevents the presumably embedded cancellation events of small-scale flux patches 
(Fig.~\ref{figure07-1}) from showing up clearly in the period prior to the flare. The splitting spot enforces the flows in 
a coherent manner over many hours, and the resulting reconnection continually adds flux to the newly formed rope. Increasing flux is known 
to eventually destabilize a flux rope \cite[e.g.,][]{2006ApJ...641..577M, 2011ApJ...734...53S, 2013arXiv1304.6981K}. 

A basically similar situation of long-lasting inflows and associated flux cancellation develops at the 
southern part of the neutral line adjacent to the satellite spots. Thus, the flows converging at the neutral line ahead of the 
leading spot, and at the satellite spots, are a prime candidate for the photospheric driver that causes the formation of a coronal flux rope 
and its subsequent destabilization. The splitting of the leading spot is a major contributor to these flows, at least in the 
northern part of the area.

The other important candidate process is flux emergence (Sect.~\ref{intro}). In the considered event, most of the flux ahead 
of the splitting spot emerges in the period of one to two days immediately before the eruption, thus forming the neutral 
line where the eruption originates. Generally, flux emergence can cause or support eruptions in five ways: (1) by reconnection with a 
pre-existing flux rope, enhancing its flux; (2) by reconnection with the ambient flux of a pre-existing flux rope, reducing the stabilizing 
effect of the ambient flux \citep{2000ApJ...545..524C}; (3) by reconnection with a pre-existing arcade field 
above the neutral line, thus forming and destabilizing a coronal flux rope \citep{2012ApJ...760...31K}; (4) by fully 
emerging a new flux rope into the corona \cite[e.g.,][]{1996SoPh..167..217L}; and (5) by partially emerging a flux 
rope into the corona and transforming the emerged part (by reconnection) into a new flux rope 
\cite[e.g.,][]{2004ApJ...610..588M, 2010A&A...514A..56A}. The observations of the event are fully consistent with the first two options, 
which both represent the destabilization of an existing flux rope, that was formed by reconnection, which in our case,
was mainly driven by the moving front half of the splitting spot. This holds true for the flux emergence in the 
arc-shaped part of the neutral line ahead of the splitting spot (i.e., under the main body of the filament), as well as for the emergence 
of the satellite sunspots and the negative flux near them (i.e., under the southern end of the filament, where the eruption starts).
The third option leaves the occurrence of the precursor flare unexplained. The fourth implies a systematic 
separation pattern of opposite-polarity flux on the sides of the neutral line as a dominant signature in the magnetogram 
\citep{2010ApJ...716L.219M}, which is not apparent in our case. The fifth is associated with strong shear flows at the neutral line, 
intrinsic to flux emergence. However, the shear flows in the present event are externally driven by the splitting spot.
Additionally, that a substantial part of the emerging flux consists of mixed small-scale polarities does not support options 
four and five. 

Finally, the very close spatial and temporal association of the weaker precursor flare with the eruptive M-class 
flare and, especially, the signs of interaction of its propagating emission front with the filament exactly at 
the onset time of the eruption are highly suggestive of a causal relationship. Since a propagating emission front 
cannot form an unstable magnetic field configuration in such a short time span, the precursor flare must play the 
role of an external trigger (a ``final drop'') acting on a configuration already close to the onset of instability.

\subsection{Splitting of the sunspot}
\label{splitting}

The manner and duration of the observed sunspot splitting is quite distinct from what is typically seen during the breakup of sunspots that 
involves one or more light bridges \citep{1987SoPh..112...49G}. The formation of light bridges is taken as evidence for convective penetration 
that disrupts the dynamic and mechanical equilibrium of a sunspot. \citet{2012ApJ...755...16L} presented an example of a regular, isolated 
sunspot splitting in this manner, and they found that the splitting required about 48 hr following the formation of a light bridge, 
although only a small part of the sunspot broke away. The splitting of the leading sunspot in this investigation is initiated by a stretching 
instead of light bridge formation, and proceeds within about a day, so much faster than usual. While the front half separates from the active 
region at a considerable speed of 210~m~s$^{-1}$ \cite[consistent with the values reported in][]{1997PASJ...49..235H}
and shows strong rotation, as well as considerable shear flows, the rear half stays essentially stationary within the active region without any 
signs of rotation or shear flows. The rotation rate of the runaway sunspot equals the highest rotation rate of a single sunspot found 
previously \citep[compare, e.g., with][]{2003SoPh..216...79B, 2008SoPh..247...39Z, 2012ApJ...754...16Y}; thus, it is a further unusual property 
of the splitting process in the considered event.

A hint at the cause of the non-standard sunspot breakup process may be found in the magnetic structure of the region. We have noted in 
Section~\ref{run} that the two fragments possess completely different magnetic connections. Essentially all flux of the separating front 
fragment connects to the neighbouring AR~11514 west of AR~11515 or to network flux in the vicinity of AR~11514. The rear fragment is apparently 
completely connected within AR~11515. Thus, we consider the possibility that the front fragment originates in the same flux system deep 
below the photosphere (or from the same rising Omega loop) as AR~11514 and becomes temporarily trapped within AR~11515 through a merger with 
the leading spot of that region, which has the same polarity. The splitting of the spot could in this case result from magnetic tension 
forces still acting on the front half. The action of such forces is also suggested by the stretching of the spot prior to the splitting. 

The alternative possibility of tension forces provided by the coronal connections to AR~11514 is unlikely for energetic reasons, as detailed 
below, and it could not cause the rotation because the coronal loops do not indicate the presence of significant twist. We therefore interpret 
the breakup and further separation of the front fragment as an attempt to merge with the active region it is magnetically connected to. 
The approach to AR~11514 is seen to continue throughout the region's rotation to the western limb. 
This scenario naturally explains the rotation of the separating fragment as the continued rise of a flux tube that is still connected to its 
source deep in the solar interior \citep{2003SoPh..216...79B, 1990SoPh..126..285V, 1994SoPh..155..301L} after the rise was temporarily stalled 
by the connection with the leading spot of AR~11515. It is also in line with the high rotation rate. The counter-clockwise rotation then implies 
a left-handed twist, and this sense of the twist is indeed indicated by the clockwise whirl of several coronal loops connected to the 
fragment (Fig.~\ref{figure08}) and by the reverse S-shaped bending of the filament's northern end (Figs.~\ref{figure03}, \ref{fig-AIA304}, 
and \ref{figure08}).

We also note that the splitting of the sunspot begins only a few hours after the nearby emergence of the satellite sunspots, 
which subsequently merge with the rear half. It is obvious that this interaction represents a major perturbation of the 
spot, and it is likely that it supports the onset of its splitting, especially if the spot is indeed in a metastable equilibrium 
of two differently rooted and connected parts.

The suggested interpretation implies that at least the front fragment of the splitting spot had not experienced a disconnection 
from its roots deep in the convection zone. The general occurrence of such a disconnection is suggested by a change in the 
dynamical properties of spots from more active to more passive evolution (with regard to the ambient convection) soon after 
their emergence \citep{1999SoPh..188..331S}; however, this is still an unresolved issue. Numerical modelling by 
\citet{2005A&A...441..337S} shows dynamical disconnection at depths of 2--6~Mm and within about three days from the 
formation of a spot. More advanced modelling \citep{2011ApJ...740...15R} instead indicates that the process would severely undermine the 
stability of the spot, if working on these scales. Therefore, dynamical disconnection should proceed only 
on time scales comparable to, or longer than, the time the spot remains coherent, and at correspondingly greater depths 
(e.g., at about 30--50~Mm for spot life times of a week). This is supported by the observational study of 
\citet{2009A&A...506..875S}, who found clear signatures of disconnection only in about a third of the investigated sunspots, with 
inferred disconnection depths in the range $\sim5\mbox{--}50$~Mm. Our interpretation of the sunspot splitting in 
AR~11515 is consistent with this more recent picture. 

Finally, we compare the kinetic energy associated with the splitting of the sunspot with the energy released by the flare. 
The kinetic energy can be calculated as $E_\mathrm{kin} = 0.5 A d \rho v_\mathrm{split}^2$, where $A$ is the area of 
a cylindrical flux tube, $d$ the depth, $\rho$ the density, and $v_\mathrm{split}$ the splitting speed of the sunspot. However, assuming that 
the proper motion of the runaway sunspot fragment at the surface is due to the vertical ascent of an inclined flux tube, not to the horizontal 
displacement of the tube, $v_\mathrm{split}$ can be replaced by $v_\mathrm{rise}$, where $v_\mathrm{rise} = v_\mathrm{split}\tan{\theta}$, 
and $\theta$ is the inclination of the tube with respect to the surface. Since the considered volume is strongly stratified, 
the area of the tube as a function of depth $z$ can be approximated as $A(z) = A_s \exp{(-z/\tau_B)}$, following flux conservation, and 
the density can be approximated as $\rho(z) = \rho_s \exp{(z/\tau_{\rho})}$. The subscript `s' refers to the value at the photosphere, while 
$\tau_B$ and $\tau_{\rho}$ correspond to the scale heights of the magnetic field strength and the density, respectively. The
kinetic energy can then be estimated as the integral $\int_{0}^{d}{0.5 A(z) \rho(z) v_\mathrm{rise}^2 dz}$. Assuming $v_\mathrm{rise}$ is 
500~m~s$^{-1}$, $\rho_s$ $10^{-7}$~g~cm$^{-3}$, $\tau_B$ 2000~km \citep{1990PhDT.......275M, 
2009ApJ...691..640R}, and $\tau_{\rho}$ 500~km \citep{2011ApJ...729....5R}, a conservative estimate of the vertical depth $d$ of 6~Mm
\citep[following][]{2005A&A...441..337S}, and using the observed spot radius of 14\farcs5 at the surface, we obtain a value of 
2$\times10^{32}$~erg for the kinetic energy. This is greater than the thermal energy of the soft X-ray 
emitting plasma and the bolometric irradiance, which are typically of the order of 10$^{30}$~erg and 10$^{31}$~erg, respectively, for 
M-class flares \citep{2012ApJ...759...71E}. This also quantitatively validates, at least to a first-order approximation, that the 
splitting of the sunspot is not the consequence but a possible driver of the flare. If disconnection happened at higher values 
of $d$, or if the runaway spot was indeed still rooted deep in the convection zone, the kinetic energy associated with the splitting
would be much higher.

A similar conclusion can be drawn regarding the relevance of the coronal flux connecting to AR~11514 for the sunspot 
splitting process. Such extended high-lying coronal flux is usually relatively close to the potential-field state, 
so it contains much less \emph{free} energy than the highly sheared core field of a flare-productive active region, 
which powers the eruptions, as in the considered event. Therefore, these loops could not store the $>10^{32}$~erg of 
free energy required to drag the front fragment of the sunspot away from AR~11515. 

Our estimate of the relation between the energy released in the considered M-class flare and the kinetic energy of the associated 
sunspot motion is the opposite of the result for an X-class flare analysed in \citet{1993SoPh..147..287A}. In that case, the whole 
sunspot closest to the flare position reached an apparent velocity of about 2~km~s$^{-1}$, but only for at most 430~sec. From 
this duration, the maximum depth of the motion was estimated to be only about 1.2~Mm, based on the Afv\'enic propagation of stresses at 
the photospheric level (caused by the flare) into the solar interior. Indeed, it is highly unlikely that a subphotospheric 
flux tube extending over many megameters would be accelerated and decelerated within the observed short time span of the sunspot 
displacement. From the much smaller depth of the motions involved, a much lower energy of $<10^{30}$~erg was estimated.
Thus, the very different time scales and implied different spatial scales of the two events of 
flare-associated sunspot changes allow for all these different conclusion.

\section{Conclusions}
\label{conclu}
This investigation of a splitting sunspot, the subsequent separation of the leading fragment from the active region, 
and the associated eruption involving a filament, M-class flare, and CME yields the following conclusions. A complex 
interplay of photospheric and coronal processes was involved in the evolution of the region to the eruption. The 
splitting of the sunspot has likely played a major role by enforcing a coherent, long-lasting inflow and shear flow at the 
photospheric neutral line ahead of the spot. This could have resulted in the formation of a coronal flux rope that held the filament and the 
eventual destabilization of the rope. The emergence of flux created the neutral line in the first place; however, we did not 
find any indications of flux rope formation directly caused by the emergence. Rather, the continued emergence has likely contributed to the 
destabilization of the flux rope. Associated with the formation of satellite sunspots, the emergence was strongest in the southern part of 
the neutral line under the filament, where the eruption started. Additionally, converging flows and resulting flux cancellation occurred at 
this part of the neutral line and have likely also contributed to flux rope formation and destabilization. A C-class precursor flare, 
possessing close spatial and temporal associations with the M-class flare and showing signs of interaction with the filament at the onset 
time of the eruption, has likely played the role of a ``final drop'' acting on an equilibrium already very close to 
the onset of instability. 

The splitting of the sunspot showed the following unusual characteristics: it began by stretching instead of light bridge formation, proceeded 
relatively fast, and set free a sunspot fragment that rotated rapidly in the course of its separation from the remaining spot, which was 
essentially stationary. The separation of the fragment from the active region was rather fast and large, although not unusual. Supported by 
the different magnetic connections of the fragments, seen as coronal loops, these properties suggest that the fragments may have originated 
in different flux systems, or Omega loops, rooted deep in the solar interior, and that the front fragment joined the flux of the active region 
only temporarily through its merging with the other flux in the leading sunspot. The perturbation of the leading spot by the emergence of 
satellite polarities shortly before and during the splitting, as well as magnetic tension forces by the subphotospheric flux connections of 
the leading part, are the factors of potential relevance for the splitting. The latter can explain the stretching of the sunspot and the strong 
separation of the front fragment, which let the fragment approach the preceding active region where its coronal magnetic connections were 
anchored. The standard interpretation of the fragment's counter-clockwise rotation as the continued rise of a twisted subphotospheric flux bundle 
(after an interruption due to trapping in the sunspot) is supported by the clockwise whirl of some coronal structures in the vicinity of the 
fragment, which indicate a left-handed twist. This interpretation of the sunspot splitting implies that at least the flux of the separating 
fragment was still connected to its roots deep in the convection zone. 

The eruption mechanism usually associated with flux dispersal and cancellation in decaying active regions gains relevance here 
in a phase of major change and additional flux emergence, owing to the motion of the split sunspot fragment towards a neighbouring neutral line. 
Previous observations of splitting sunspots found flaring associated with a very similar driving, often including a major flare 
\citep{1991ApJ...380..282W, 1994SoPh..150..199S, 1998ApJ...502..493D}. A similar situation occurs when sunspots of different active regions, 
or within complex regions, approach each other, typically also causing major eruptions 
\citep[e.g.,][]{1986AdSpR...6...29K, 2007PASJ...59S.779K, 2012ApJ...748...77S}. Thus, events like the one studied here may serve as testbeds 
for detailed numerical studies of this important eruption mechanism under relatively well defined observational constraints. In addition to 
the comprehensive observational coverage provided by current instruments, the measurements of the magnetic and velocity fields are relatively 
reliable, thanks to their coherent and high values. This may also allow checking the key conjecture of this mechanism---that a flux rope forms 
prior to the eruption---against the opposing view of eruption onset in a sufficiently sheared arcade \citep{2012ApJ...760...81K}.

Multi-wavelength observations with good spatial and temporal resolution, including spectroscopy and polarimetry in the 
photosphere and chromosphere, have proven to be crucial in this study for resolving the complex photospheric dynamics 
and for at least partially disentangling their equally complex relationships to the resulting coronal evolution. 
These diagnostic tools provide important information on the nature and dynamics of plasma and magnetic 
fields at the site of the activity. It is anticipated that new instruments such as the GREGOR Fabry P\'erot 
Interferometer \citep[GFPI,][and references therein]{2012SPIE.8446E..79P,2012AN....333..880P}, 
the BLue Imaging Solar Spectrometer \citep[BLISS,][]{2010AN....331..648D,2012SPIE.8446E..79P,2013OptEn..52h1606P},
and the Chromospheric Magnetometer \citep[ChroMag,][]{2012AAS...22013506B} will be pivotal for carrying out 
such investigations in even more detail in the future.

\begin{acknowledgements}
HMI data are courtesy of NASA/SDO and the HMI science team. They are provided by the Joint Science Operations 
Center -- Science Data Processing at Stanford University. The GOES X-ray flux measurements were made available 
by the National Geophysical Data Center. The SOHO/LASCO CME catalogue is maintained by the CDAW Data Center.
The Vacuum Tower Telescope and ChroTel are operated by the Kiepenheuer-Institute 
for Solar Physics in Freiburg, Germany, at the Spanish Observatorio del Teide, Tenerife, Canary Islands. The 
ChroTel filtergraph has been developed by the Kiepenheuer-Institute in cooperation with the High Altitude 
Observatory in Boulder, CO, USA. In particular, we would like to thank Dr. Thomas Kentischer, Clemens 
Halbgewachs, and Hans-Peter Doerr for their efforts to improve ChroTel and to ensure high-quality data. R.E.L. 
and C.D. were supported by grant DE 787/3-1 of the German Science Foundation (DFG). 
B.K. acknowledges the hospitality of the solar group at the Yunnan Astronomical Observatory, where most of 
his work was carried out, and the associated support by the Chinese Academy of Sciences under grant no.~2012T1J0017. 
He also acknowledges support by the DFG and the STFC. We are thankful to 
Drs. Christian Beck, Christian Bethge, and Christoph Kuckein for carefully reading the manuscript and 
providing valuable suggestions. We appreciate the inputs provided by Dr. Alexander Warmuth and Dr. Haimin Wang. 
We thank the anonymous referee and the editor for their helpful comments and very constructive suggestions.
\end{acknowledgements}

\Online
\begin{appendix}
\section{Temporal evolution of AIA loops}
\end{appendix}

\end{document}